\documentclass[useAMS,usenatbib]{mn2e}
\usepackage{graphicx}
\usepackage[T1]{fontenc}
\usepackage{aecompl}
\title[jet-disc connection in Fermi blazars]{Core-dominance parameter, black hole mass and jet-disc connection in Fermi blazars}
\author[Y.Y. Chen, X. Zhang, H.J. Zhang and X.L. Yu]{Y.Y. Chen, X. Zhang, H.J. Zhang\thanks{E-mail:kmzhanghj@163.com} and X.L. Yu\\
Department of physics, Yunnan Normal University, Kunming 650500, China\\}
\begin{document}

\maketitle

\label{firstpage}

\begin{abstract}
We study the relationship between jet power and accretion for Fermi and non-Fermi blazars, respectively. We also compare the relevant parameter between them. Our main results are as follows. (i) Fermi and non-Fermi blazars have significant difference in redshift, black hole mass, and broad line luminosity. (ii) Fermi blazars have higher average core-dominance parameter than non-Fermi blazars, which suggests that Fermi blazars have strong beaming effect. (iii) We find significant correlation between broad line emission and jet power for Fermi and non-Fermi blazars, respectively, which suggests a direct tight connection between jet and accretion. (iv) The accretion and black hole mass may have a different contribution to jet power for Fermi and non-Fermi blazars, respectively.
\end{abstract}


\section{Introduction}
Blazars are the most extreme classes of active galactic nuclei (AGN), showing large amplitude and rapid variability, superluminal motion, and strong emission. Blazars also host a jet, pointing almost directly to the observer (Urry \& Padovani 1995). Their extreme observation properties can be explained by a beaming effect. Because of a relativistic beaming, the emission that is dominated by a relativistic jet is highly boosted in the line of observer's sight (Urry \& Padovani 1995). According to emission line features, blazars are often divided into Flat Spectrum Radio Quasars (FSRQs) and BL Lacerate objects (BL Lacs). Some blazars with an equivalent width (EW) of the emission lines in the rest frame EW$>5\rm{\AA}$ are classified as FSRQs (e.g., Scarpa \& Falomo 1997; Urry \& Padovani 1995). However, Ghisellini et al. (2011) have suggested that this classification is not reliable, and more over does not reflect any intrinsic property or difference within the blazars class. Therefore they introduced a more physical classification based on the luminosity of the broad emission lines measured in Eddington units, and the divided line is of the order of $\rm{L_{BLR}/L_{Edd}\sim5\times10^{-4}}$. Sbarrato et al. (2012) and Xiong et al. (2014) have confirmed this result. Giommi et al. (2012, 2013) have also deeply investigated the unreliability of the EW classification. They suggested that blazars should be divided into high- and low-ionization sources. Landt et al. (2004) have also introduced an analogous classification criterion. They found that it is possible to discriminate between objects with intrinsically weak or strong narrow emission lines by studying the [$\rm{O_{II}}$] and [$\rm{O_{III}}$] EW plane.

Since the launch of the Fermi satellite, we have entered in a new era of blazars research (Abdo et al. 2009, 2010). Up to now, the Large Area Telescope (LAT) has detected hundreds of blazars because it has about 20 flods better sensitivity than its predecessor EGRET in the 0.1-100 Gev energy range. However, at present, there is outstanding question about the AGN, which is unclear that ``why are some sources $\rm{\gamma}$-ray loud and others are $\rm{\gamma}$-ray quiet?''. Many answers have been proposed to explain this question, such as Doppler boosting, apparent jet speed, apparent opening angle, VLBI core flux densities and brightness temperatures. Blazars detected by LAT are more likely to have higher Doppler factors (e.g, Lister et al. 2009; Savolainen et al. 2010; Tornikoski et al. 2010) and larger apparent opening angles (e.g., Pushkarev et al. 2009) than those not detected by LAT. Many authors have suggested a close connection between the $\rm{\gamma}$-ray emission and radio properties of AGN. Kovalev et al. (2009) have suggested that LAT-detected blazars are brighter and more luminous in the radio domain at parsec scales. Lister et al. (2009) also suggested that LAT-detected blazars have high apparent jet speeds. Pushkarev et al. (2012) also showed that the Fermi AGNs have higher VLBI core flux densities and brightness temperatures. Ghisellini et al. (2010) studied the general physical properties of bright Fermi blazars. According to the SEDs, they got the jet power and disk luminosity. They found a positive correlation between jet power and the luminosity of the accretion disc in those blazars. Xiong and Zhang (2014) also studied the physical properties of Fermi blazars. Sbarrato et al. (2014) found a positive relation between radio luminosity and broad line luminosity in AGNs. However, it is unclear whether there is difference in jet and accretion for LAT blazars and non-LAT blazars. Therefore, we tried to study this question.

In this paper, we collect a large sample of LAT-detected and non-LAT detected blazars, and study the properties of Fermi blazars. The main results of our analysis concern the relation between the jet power and accretion, the relation between jet power and black hole mass in Fermi and no-Fermi blazars, respectively. The paper is structured as follows: we present the sample in Sect.2; the results are in Sect.3; the discussions in Sect.4; the conclusions are in Sect.5. A $\Lambda$CDM cosmology with $H_{\rm{0}}={\rm{70Km s^{-1} Mpc^{-1}}}$, $\Omega_{\rm{m}}=0.27$, $\Omega_{\rm{\Lambda}}=0.73$ is adopted. The energy spectral index $\rm{\alpha}$ is defined such that $\rm{S_{\nu}\propto\nu^{-\alpha}}$

\section{ THE SAMPLE}
The major selection criteria for the sample that we tried to use radio catalogues to get the widest possible sample of blazars from their radio properties, and then split them into Fermi detected sources and non-Fermi detections. Massaro et al.(2009) have described ``Multifrequency Catalogue of BLAZARS'', also named Roma-BZCAT. The Roma-BZCAT contains the lists of blazars, which are classified in three main groups based on their spectral properties. Each blazar is identified by a three-letter code. The codes are respectively BZB: BL Lac objects, used for AGNs with a featureless optical spectrum or only with absorption lines of galaxian origin and weak and narrow emission lines; BZQ: flat-spectrum radio quasars, with an optical spectrum showing broad emission lines and dominant blazar characteristics; BZU: blazars of uncertain type, adopted for sources with peculiar characteristics but also showing blazar activity. The widest possible sample of blazars also were included in BZCAT (Massaro et al.2009: The Roma BZCAT) and have reliable radio core and extended luminosity at 1.4 GHz, redshift, black hole mass and broad line luminosity (used as a tracer of the accretion).

Firstly, we consider the following samples of blazars to get the radio core and extended luminoaity at 1.4 GHz: Kharb et al. (2010), Antonucci \& Ulvestad (1985), Cassaro et al. (1999), Murphy et al. (1993), Landt et al. (2008), Caccianiga et al. (2004), Giroletti et al. (2004). Secondly, we consider the following samples of blazars to get the broad line data: Celotti et al. (1997), Cao \& Jiang (1999), Wang et al. (2002, 2004), Liu et al. (2006), Xie et al. (2007), Sbarrato et al. (2012), Chai et al. (2012), Shen et al. (2011), Shaw et al. (2012). Thirdly, we consider the following samples of blazars to get black hole mass: Woo \& Urry (2002), Cao et al. (2002), D'Elia et al. (2003), Liang \& Liu (2003), Xie et al. (2004), Liu et al. (2006), Fan et al. (2008), Zhou \& Cao (2009), Xu et al. (2009), Wu et al. (2008), Ghisellini et al. (2011), Zhang et al. (2012), Sbarrato et al. (2012), Chai et al. (2012), Leon-Tavares et al. (2011a), Shen et al. (2011), Shaw et al. (2012). At last, we cross-correlated these sample with clean blazars detected by Fermi LAT in two years of scientific operation (Abdo et al. 2012, 2FGL; Ackermann et al. 2011a, 2LAC). In total, we have a sample containing 177 clean Fermi blazars (96 Fermi FSRQs and 81 Fermi BL Lacs) and 133 non Fermi blazars (105 non-Fermi FSRQs and 28 non-Fermi BL Lacs).

We also note that there may have a select bias about our samples, because our samples only contain the 2LAC clean Fermi blazars and non-EGRET detected blazars in our non-Fermi blazars sample. And all blazars are BZQ or BZB in our sample. The BZU is not contained in our sample. But we find that the redshift distributions of our sample are agree with the Rom-BZCAT. Therefore the select bias should not have large influence for our main results in a certain extent. Xiong and Zhang (2014) have described in detail the calculation of black hole mass and broad line luminosity. In order to reduce the uncertainty, we tried to select the data from a same paper and /or a uniform method as soon as possible. Tremaine et al. (2002) have suggested that the uncertainty in the $\rm{M_{BH}-\sigma}$ relation is small, $\leq0.21$ dex; and the uncertainty on the zero point of the line width-luminosity-mass relation is approximately 0.5 dex (Gebhardt et al. 2000; Ferrarese et al. 2001); MuLure \& Dunlop (2001) have suggested that the $\rm{M_{BH}-M_{R}}$ correlation for quasars host galaxies has an uncertainty of 0.6 dex (Wang et al. 2004). According to set the Ly$\rm{\alpha}$ flux contribution to 100, and the relative weights of the H$\rm{\alpha}$, H$\rm{\beta}$, $\rm{Mg_{II}}$ and $\rm{C_{IV}}$ lines to 77, 22, 34 and 63, respectively (see Francis et al. 1991), Celotti, Padovani, \& Ghisellini (1997) have calculated the broad line luminosities. We follow Celotti, Padovani, \& Ghisellini (1997) and calculate the broad line luminosity for our sample. When more than one line is presented, we calculate the simple average of broad line luminosity estimated from each line. We assume that the uncertainty of broad line luminosity is 0.5dex. Analogously, when more than one black hole mass is gotten, we calculate average black hole mass.

The core-dominance parameter that the ratio of the beamed radio core flux density ($\rm{S_{core}}$) to the unbeamed extend radio flux density ($\rm{S_{ext}}$) has routinely been used as a statistical indicator of Doppler beaming and orientation (Orr \& Browne 1982; Kapahi \& Saikia 1982; Kharb \& Shastri 2004). We have made a K-correction for the observed flux by using $\rm{S(\nu)}$=$\rm{S_{\nu}^{ob}(1+z)^{\alpha-1}}$. The luminosity is calculated from the relation $\rm{L_{\nu}}$=$\rm{{4\pi}d_{L}^{2}S_{\nu}}$. We calculate the core-dominance parameter ($\rm{R_{c}}$=$\rm{\frac{S_{core}}{S_{ext}}(1+z)^{\alpha_{core}-{\alpha_{ext}}}}$, with $\rm{\alpha_{core}}=0$, $\rm{\alpha_{ext}}=0.8$)

The jet power also can be derived from the lobe low frequency radio emission under the assumption of minimum energy arguments(e.g., Rawlings \& Saunders 1991; Willott et al. 1999). This approach now is widely used to estimate the
jet kinetic energy in AGNs. Meyer et al.(2011) used the following formula to estimated the cavity kinetic power,i.e.,
\begin{equation}
\rm{\log{P_{cav}}}=\rm{0.64(\pm0.09)(\log{L_{300}}-40)+43.54(\pm0.12)}
\end{equation}
where $\rm{L_{300}}$ is the extend luminosity at 300 MHz, the unit of jet power is erg $\rm{s^{-1}}$, which is continuous over $\sim6-8$ decades in $\rm{P_{jet}}$ and $\rm{P_{radio}}$ with a scatter of $\sim0.7dex$ and $\rm{P_{cav}=P_{jet}}$. We extrapolate the extend 1.4 GHz flux density to calculate the extend 300 MHz flux density, by assuming a spectral index of $\rm{\alpha=1.2}$ in this paper. We use equation (1) to get the jet power.

The relevant data for Fermi blazars is listed in Table 2 with the following headings: column (1) the name of the Fermi blazars (2FGL); column (2) classification of Fermi blazars (BZQ=FSRQ, BZB=BL Lac); column (3) the redshift; column (4) the radio core flux density at 1.4 GHz, the units is Jy; column (5) the radio extended flux density at 1.4 GHz, the units is mJy; column (6) the references of column (4) and column (5); column (7) the black hole mass; column (8) the references of the black hole mass; column (9) the broad line luminosity, the units is erg$s^{-1}$; column (10) the references of broad line luminosity. The relevant data for non-Fermi blazars is also listed in Table 3.

\section{RESULTS}
\subsection{The distributions}
We make the histogram about redshift for the various classes in Figs. 1. From Fig.2 of Rom-BZCAT, the redshift distributions of BL Lacs are much closer than that of FSRQs and their distribution peaks at z$\cong$0.3, whereas FSRQs show a broad maximum between 0.6 and 1.5. There are only very few BL Lacs at redshift higher than 0.8. So our results agree with the results of Rom-BZCAT in the redshift distributions. From Figs.1, we can see that the range of redshift is $0<z<2.5$ and $0<z<3.5$ for Fermi and non-Fermi blazars. The mean redshift are listed in Table 1. Through nonparametric Kolmogorov-Smirnov (K-S) test, we get that the distributions of redshift between all Fermi blazars and all non-Fermi blazars, between Fermi FSRQs and non-Fermi FSRQs are significantly different (see Table 1, significant probability P$<$0.05). However, there is no significant difference between Fermi BL Lacs and non-Fermi BL Lacs. The non-Fermi FSRQs have higher average redshift than Fermi FSRQs. The Fermi BL Lacs have higher average redshift than non-Fermi BL Lacs. Linford et al. (2011) suggested that it is still enough to rule out redshift as the cause of LAT non-detection. If assuming the non-LAT BL Lacs might have been too far away to detect their $\rm{\gamma}$-rays. We find that the LAT BL Lacs have higher average redshift than non-LAT BL Lacs. Therefore we confirm the result of Linford et al. (2011). They also found that there is no strong correlation between redshift and $\rm{\gamma}$-ray flux for the BL Lac objects.
\begin{figure}
\includegraphics[width=8.5cm,height=7.5cm]{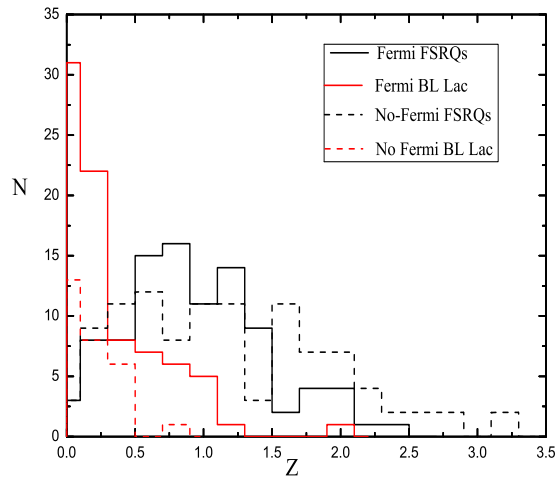}
\vspace{0pc}
\caption{Distributions of redshifts of Fermi and non-Fermi blazars. The black solid line is Fermi FSRQs. The red solid line is Fermi BL Lacs. The black dashed line is non-Fermi FSRQs. The red dashed line is non-Fermi BL Lacs.}
\label{sample-fig}
\end{figure}
\begin{figure}
\includegraphics[width=8.5cm,height=7.5cm]{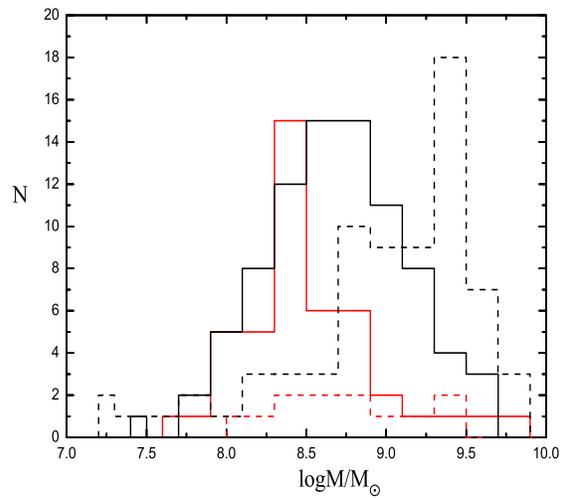}
\vspace{0pc}
\caption{Distributions of black hole mass of Fermi and non-Fermi blazars. The meanings of different lines are as same as Fig.1. }
\label{sample-fig}
\end{figure}
\begin{figure}
\includegraphics[width=8.5cm,height=7.5cm]{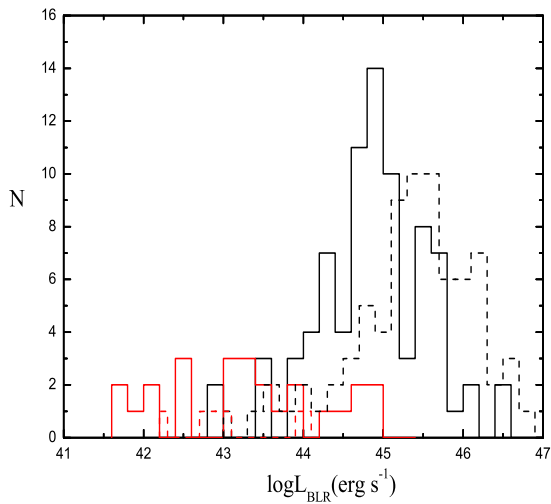}
\vspace{0pc}
\caption{Distributions of broad line luminosity of Fermi and non-Fermi blazars. The meanings of different lines are as same as Fig.1.}
\label{sample-fig}
\end{figure}

The black hole mass distributions of the various classes are shown in Figs. 2. The mean black hole mass are listed in Table 1 for various classes. Through the K-S test, we get that the distributions of black hole mass between all Fermi blazars and all non-Fermi blazars, between Fermi FSRQs and non-Fermi FSRQs are significantly different. However, there is no significant difference between Fermi BL Lacs and non-Fermi BL Lacs (see Table 1). Compared with Fermi FSRQs, the non-Fermi FSRQs have higher mean black hole mass. Compared with the Fermi BL Lacs, the non-Fermi BL Lacs have higher mean black hole mass. There may be a general think that the Fermi blazars may have large black hole mass (Ghisellini et al. 2010). We should notice that the $\rm{\gamma}$-ray narrow line Seyfert 1 have lower black hole mass than blazars, whereas it can be detected by LAT. Meier (1999) have suggested that it is not necessary to have a relatively massive black hole to produce powerful jet. According to the current accretion and jet production theory (Blandford \& Znajek 1977; Meier 1999; Xie et al. 2006, 2007; Chai et al. 2012), jet power is linked with the spinning of black hole.
\begin{figure}
\includegraphics[width=8.5cm,height=7.5cm]{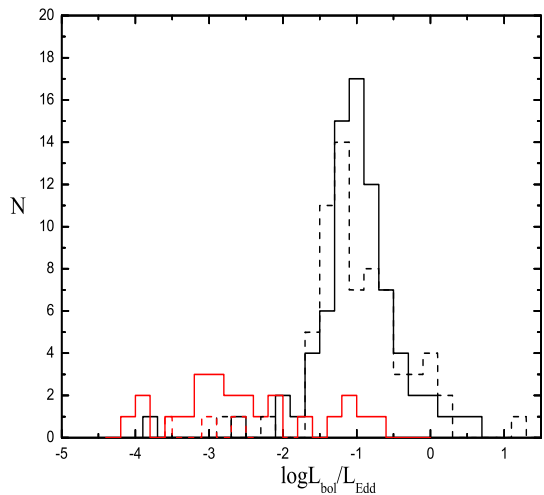}
\vspace{0pc}
\caption{Distributions of Eddington ratios of Fermi and non-Fermi blazars. The meanings of different lines are as same as Fig.1.}
\label{sample-fig}
\end{figure}
\begin{figure}
\includegraphics[width=8.5cm,height=7.5cm]{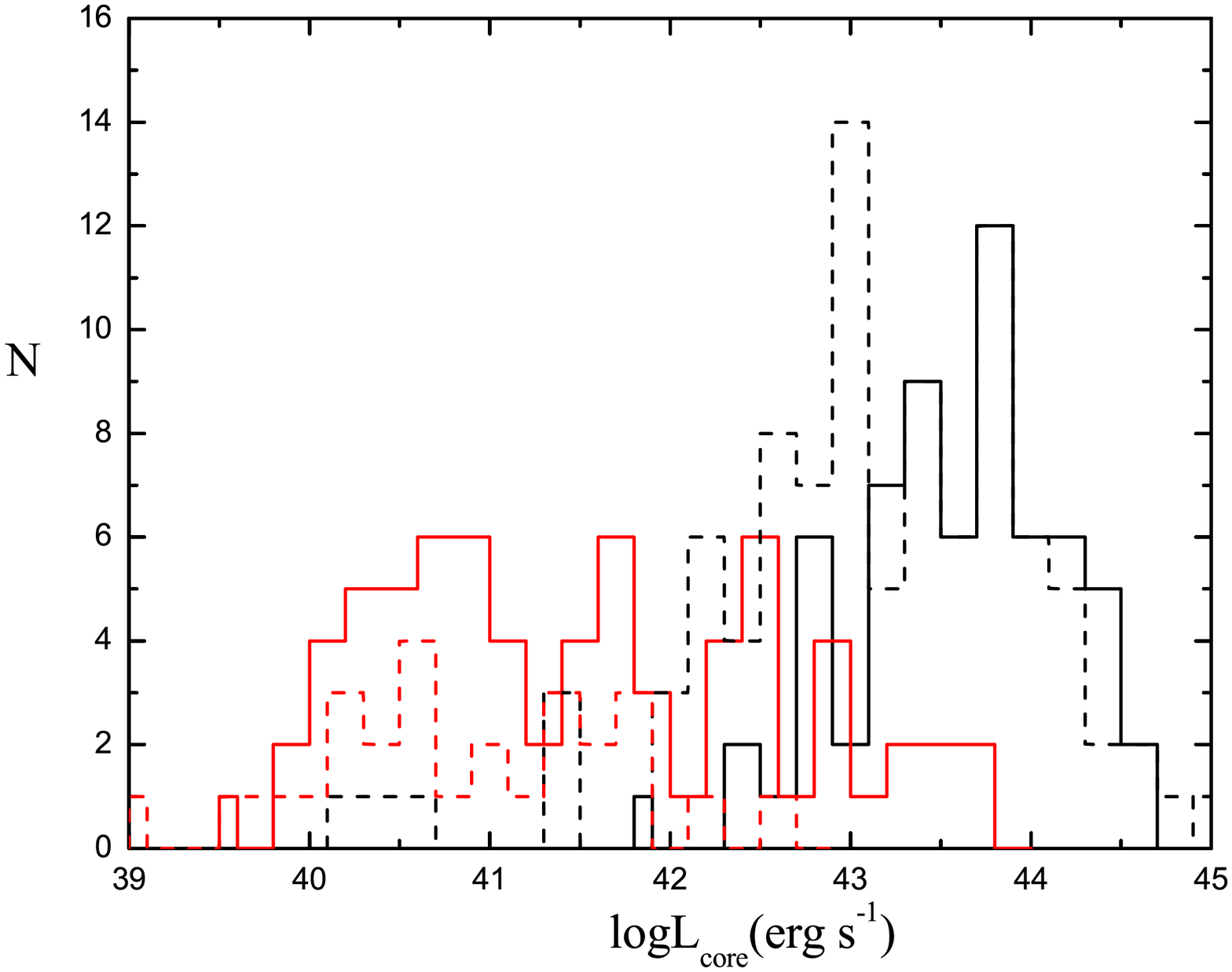}
\vspace{0pc}
\caption{Distributions of core luminosity of Fermi and non-Fermi blazars. The meanings of different lines are as same as Fig.1.}
\label{sample-fig}
\end{figure}

The broad line luminosity distributions for the various classes are shown in Figs. 3. The mean broad line luminosity are listed in Table 1 for various classes. Through the K-S test, we get that the broad line luminosity distributions between all Fermi blazars and non-Fermi blazars, between Fermi FSRQs and non-Fermi FSRQs are significantly different (see Table 1). Compared with Fermi FSRQs, the non-Fermi FSRQs have higher mean broad line luminosity. The Eddington ratio distributions for the various classes are shown in Figs. 4 ($\rm{L_{bol}/L_{Edd}}$, $\rm{L_{bol}}$$\approx\rm{10L_{BLR}}$, $\rm{L_{Edd}}$=$\rm{1.3\times10^{38}(M/M_{\odot})erg s^{-1}}$). The mean Eddington ratios are also listed in Table 1. Through the K-S test, we get that the Eddington ratio distributions between all Fermi blazars and non-Fermi blazars, between Fermi FSRQs and non-Fermi FSRQs have no significant difference (see Table 1). Compared with Fermi FSRQs, the non-Fermi FSRQs have higher mean Eddington ratios. Due to few non-Fermi BL Lacs having broad line luminosity, we only compare broad line luminosity distributions between Fermi FSRQs and non-Fermi FSRQs. Ghisellini et al. (1998) have suggested that the difference between the jet production mechanisms may also manifest in the observed luminosity, which has been proposed to unify the subclasses of blazars as a blazar sequence. Ghisellini et al. (2009a, 2010) have suggested that the difference between BL Lacs and FSRQs may be associated with the different accretion rate, because of a very weak BLR may form if the accretion rate is low than $10^{-2}\rm{L_{Edd}}$ (Ho 2008). Therefore, the BLR is
also related to the accretion disk structure and the disk radiative efficiency. The division between BL Lacs and FSRQs may be observationally controlled by the luminosity of the BLR measured in Eddington units (Ghisellini et al. 2011; Sbarrato et al. 2012; Zhang et al. 2014).
\begin{table*}
\begin{minipage}{155mm}
\centering
\caption{The KS test on properties of Fermi and non-Fermi blazars.}
\begin{tabular}{@{}crcccccccccccccccrl@{}}
\hline\hline
Parameter & probability & Significantly different &   subsets      &   mean(Fermi)   &   mean(non-Fermi)\\
\hline
z         &   0.002     &       YES               &   allblazars   &    $0.74\pm0.56$ &   $1.06\pm0.79$ \\
          &   0.011     &       YES               &   FSRQs        &    $1.03\pm0.52$ &   $1.27\pm0.75$ \\
          &   0.249     &       NO                &   BL           &    $0.40\pm0.37$ &   $0.28\pm0.19$ \\
$\rm{\log{(M/M_{\odot})}}$ & $2.36\times10^{-6}$ &  YES & allblazars & $8.72\pm0.44$  &   $9.03\pm0.59$   \\
          & $3.28\times10^{-5}$ &  YES  &  FSRQs &  $8.78\pm0.43$ &   $9.07\pm0.61$ \\
          & 0.408       &    NO   &   BL   &   $8.61\pm0.43$     &  $8.81\pm0.43$  \\
$\rm{\log{L_{BLR}}}$ & $1.56\times10^{-7}$ & YES & allblazars &  $44.60\pm1.03$  &  $45.33\pm0.94$  \\
                     & $1.5\times10^{-5}$  & YES &   FSRQs    &  $44.97\pm0.70$  &  $45.44\pm0.79$  \\
$\rm{\log{L_{bol}/L_{Edd}}}$ & 0.084 &  NO & allblazars &  $-1.26\pm0.96$  &  $-0.93\pm0.73$   \\
                             & 0.408 & NO  & FSRQs &  $-0.90\pm0.60$  &  $-0.84\pm0.60$  \\
$\rm{\log{L_{core}}}$ & 0.435 & NO &  allblazars &   $42.59\pm1.36$   &  $42.71\pm1.27$  \\
                      & 0.001 & YES &  FSRQs &   $43.67\pm0.58$   &  $43.20\pm0.88$  \\
                      & 0.071 & NO &  BL &   $41.60\pm1.09$   &  $40.98\pm0.84$  \\
$\rm{\log{L_{ext}}}$ & 0.086 & NO &  allblazars &   $41.52\pm1.30$   &  $41.65\pm1.15$  \\
                      & $2.75\times10^{-4} $ & YES &  FSRQs &   $42.50\pm0.76$   &  $41.98\pm0.99$  \\
                      & 0.672 & NO &  BL &   $40.71\pm1.08$   &  $40.51\pm0.92$  \\
$\rm{\log{R_{c}}}$ & 0.50 & NO &  allblazars &   $1.10\pm0.68$   &  $1.04\pm0.81$  \\
                      & 0.952 & NO &  FSRQs &   $1.18\pm0.57$   &  $1.22\pm0.73$  \\
                      & 0.004 & YES &  BL &   $1.03\pm0.77$   &  $0.40\pm0.79$  \\
\hline
\end{tabular}
\end{minipage}
\end{table*}

The core luminosity distributions for the various classes are shown in Figs. 5. The mean core luminosities are listed in Table 1 for various classes. Through the K-S test, we get that the core luminosity distributions between all Fermi blazars and all non-Fermi blazars, between Fermi BL Lacs and non-Fermi BL Lacs have no significant difference (see Table 1). However, there is significant difference between Fermi FSRQs and non-Fermi FSRQs. The Fermi FSRQs have higher mean core luminosity than non-Fermi FSRQs. The Fermi BL Lacs have higher mean core luminosity than non-Fermi BL Lacs.
\begin{figure}
\includegraphics[width=8.5cm,height=7.5cm]{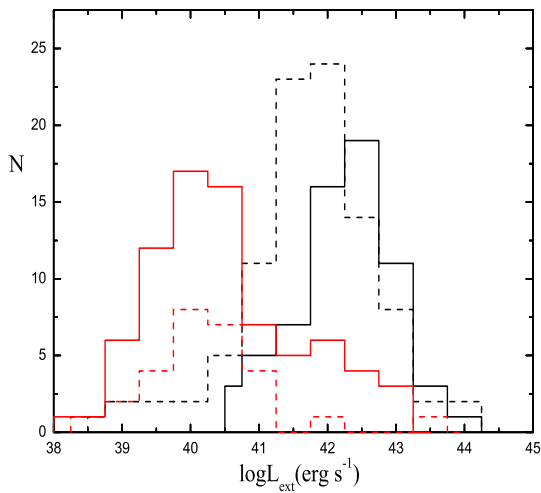}
\vspace{0pc}
\caption{Distributions of extended luminosity of Fermi and non-Fermi blazars. The meanings of different lines are as same as Fig.1.}
\label{sample-fig}
\end{figure}
\begin{figure}
\includegraphics[width=8.5cm,height=7.5cm]{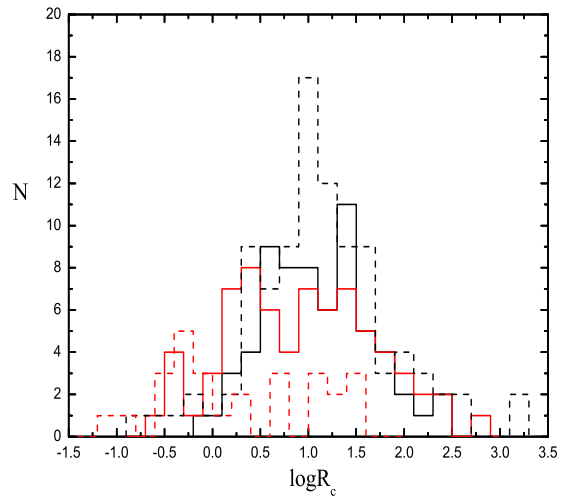}
\vspace{0pc}
\caption{Distributions of core-dominance parameter of Fermi and non-Fermi blazars. The meanings of different lines are as same as Fig.1.}
\label{sample-fig}
\end{figure}

The extended radio luminosity distributions for the various classes are shown in Figs. 6. The mean extended radio luminosities are listed in Table 1 for various classes. Through the K-S test, we get that the extended radio luminosity between all Fermi blazars and all non-Fermi blazars, between Fermi BL Lacs and non-Fermi BL Lacs have no significant difference. However, there is significant difference between Fermi FSRQs and non-Fermi FSRQs (see Table 1). The extend radio luminosity can be used to indicate the intrinsic jet power. This result may suggest that there have no significant different in intrinsic jet power for all Fermi and all non-Fermi blazars. Compared with Fermi FSRQs, the non-Fermi FSRQs have lower average extended radio luminosity. Compared with non-Fermi BL Lacs, the Fermi BL Lacs have higher average extended radio luminosity.
\begin{figure}
\includegraphics[width=9.5cm,height=12cm]{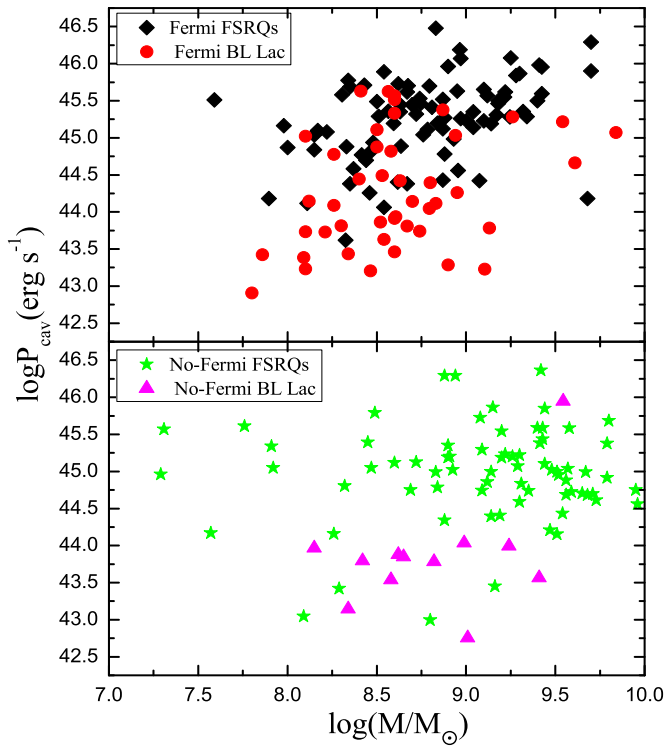}
\vspace{0pc}
\caption{The jet power as a function of black hole mass for Fermi (top) and non-Fermi blazars (bottom). The black square is Fermi FSRQs . The filled red circle is Fermi BL Lacs. The green stars is non-Fermi FSRQs. The magenta triangles is non-Fermi BL Lacs.}
\label{sample-fig}
\end{figure}

The core-dominance parameter distributions for the various classes are shown in Figs.7. The range of core-dominance parameter is from $10^{-0.75}$ to $10^{3.0}$ for all Fermi blazars; the scope of core-dominance parameter is from $10^{-1.25}$ to $10^{3.50}$ for all non-Fermi blazars. The average core-dominance parameters are listed in Table 1 for various classes. Through the K-S test, we get that the distributions between all Fermi blazars and all non-Fermi blazars, between Fermi FSRQs and non-Fermi FSRQs have no significant difference. However, there is significant difference between Fermi BL Lacs and non-Fermi BL Lacs (see Table 1). Compared with all non-Fermi blazars, the all Fermi blazars have higher average core-dominance parameter, which suggests that the Fermi blazars have strong beaming effect. The Fermi FSRQs have lower average core-dominance parameter than non-Fermi FSRQs. The Fermi BL Lacs have significantly higher average core-dominance parameter than non-Fermi BL Lacs. Kharb et al. (2010) found that the ratio of the radio core luminosity to the k-corrected optical luminosity ($\rm{\log R_{\nu}}$=$\rm{\log\frac{L_{core}}{L_{opt}}}$=$\rm{(\log L_{core}+M_{abs}/2.5)-13.7}$) appears to be a better indicator of orientation than traditionally used radio core-dominance parameter ($\rm{R_{c}}$). They suggested that the extended radio luminosity may be affected by interaction with the environment on Kiloparsec-scales. We find that there is no significant difference between Fermi and non-Fermi FSRQs in the distributions of core-dominance parameter. Because of the FSRQs have a rich dense environment, which may lead to above result that the distributions difference of core-dominance parameter between Fermi and non-Fermi FSRQs.

\begin{figure}
\includegraphics[width=9.5cm,height=12cm]{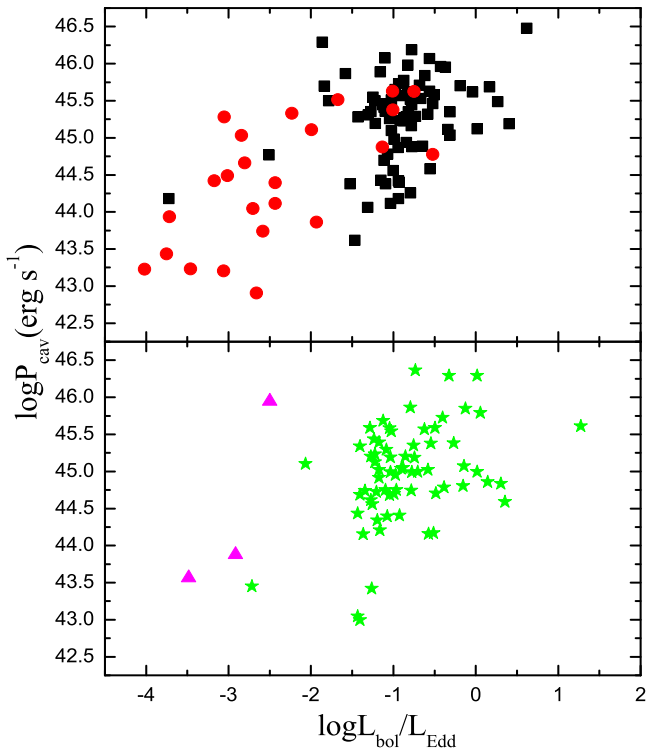}
\vspace{0pc}
\caption{The jet power as a function of Eddington ratio of Fermi (top) and non-Fermi (bottom) blazars. The meanings of different symbols are as same as Fig.8.}
\label{sample-fig}
\end{figure}
\begin{figure}
\includegraphics[width=8.5cm,height=7.5cm]{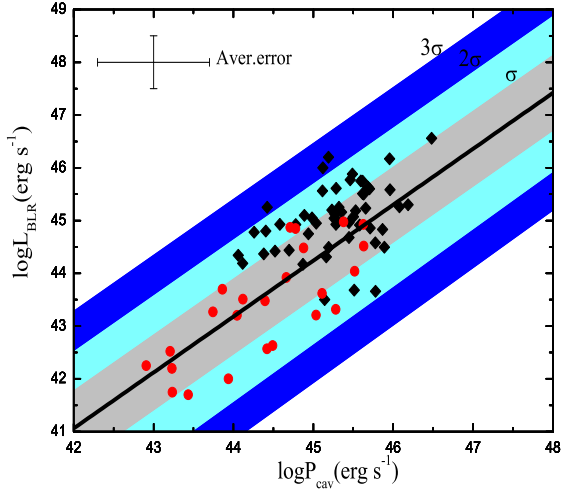}
\vspace{0pc}
\caption{The broad line luminosity as a function of jet power of Fermi blazars. Shaded colored areas correspond to 1, 2
and 3 $\rm{\sigma}$ (vertical) dispersion, $\rm{\sigma=0.74}$. The black line is the best least square fit ($\rm{\log L_{BLR}=1.06\log P_{cav}-3.46}$). The average error bar corresponds to an uncertainty of
a factor 0.5 in $\rm{\log L_{BLR}}$ and 0.7 in $\rm{\log P_{cav}}$. The meanings of different symbols are as same as Fig.8.}
\label{sample-fig}
\end{figure}
\begin{figure}
\includegraphics[width=8.5cm,height=7.5cm]{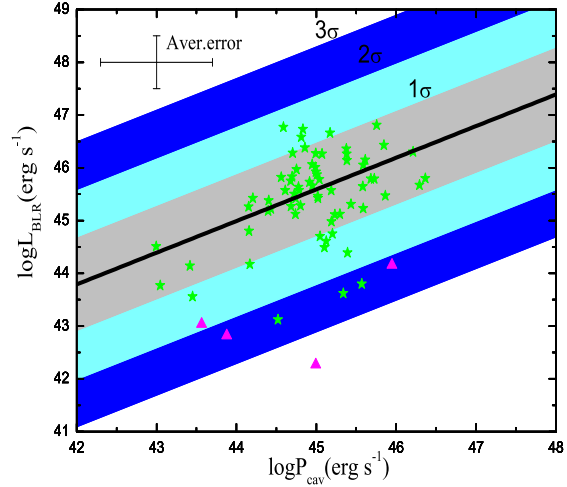}
\vspace{0pc}
\caption{The broad line luminosity as a function of jet power of non-Fermi blazars. Shaded colored areas correspond to 1, 2
and 3 $\rm{\sigma}$ (vertical) dispersion, $\rm{\sigma=0.90}$. The black line is the best least square fit ($\rm{\log L_{BLR}=0.59\log P_{cav}+18.59}$). The meanings of different symbols are as same as Fig.8.}
\label{sample-fig}
\end{figure}
\subsection{Jet power vs black hole mass and Eddington ratio}
The relationship between jet power and black hole mass is shown in Figs.8. Different symbols correspond to blazars of different classes. We use the Pearson's analysis to analyze the correlations between jet power and black hole mass for all blazars (Ackermann et al. 2011b; Padovani 1992; Machalski \& Jamrozy 2006). We find significant correlations between jet power and black hole mass for Fermi blazars (number of points N=129, significance level P$<$0.0001, coefficient of correlation r=0.40). However, there have no significant correlations for non-Fermi blazar( N=84, P=0.11, r=0.18). Figs.9 shows the relationship between jet power and Eddington ratio. We also find that there are significant correlations between jet power and Eddington ratio for both Fermi blazars and non-Fermi blazars (Fermi blazars: N=101, P$<$0.0001, r=0.63; non-Fermi blazars: N=70, P$<$0.0001, r=0.42).
\begin{figure}
\includegraphics[width=9cm,height=15cm]{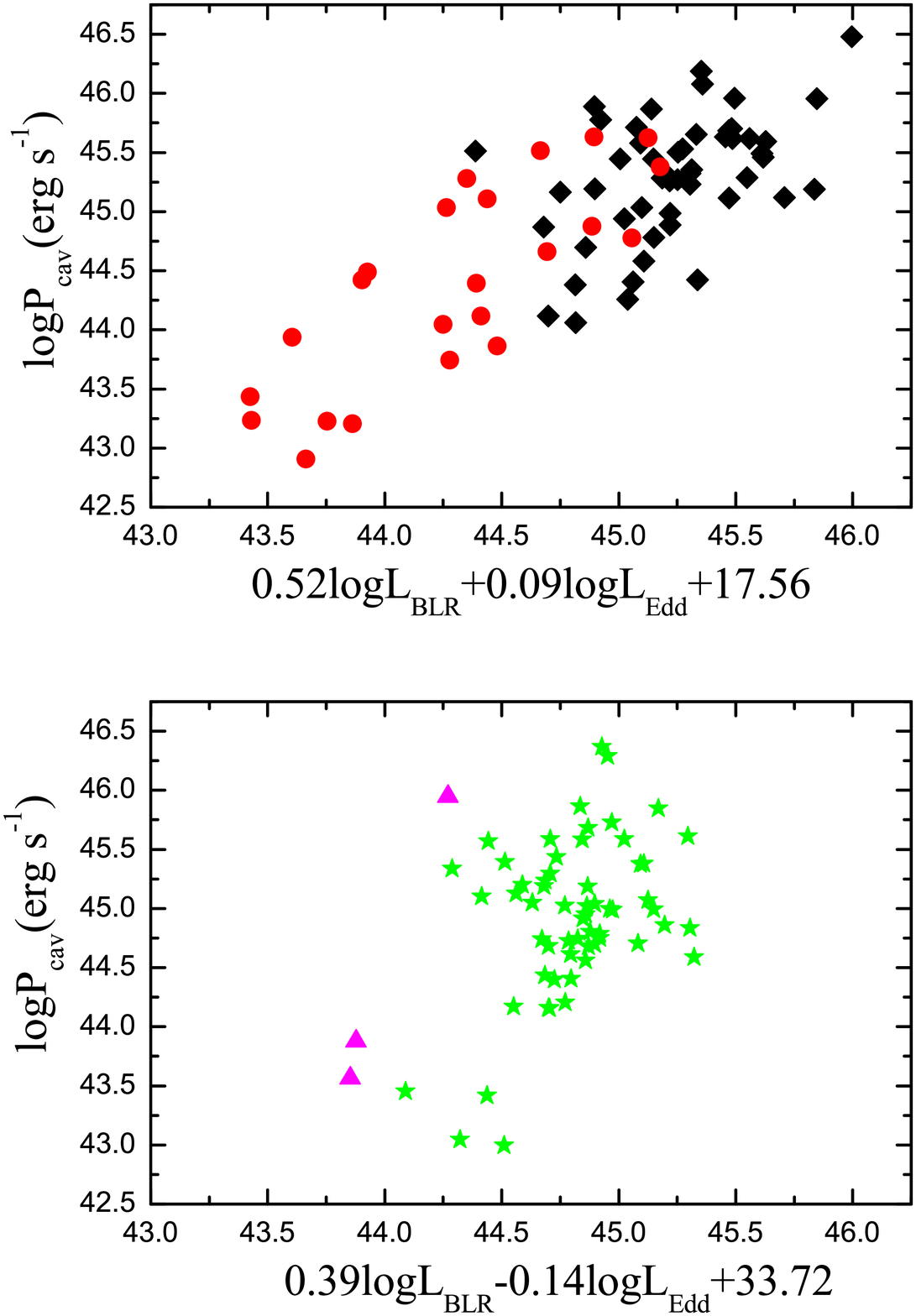}
\vspace{0pc}
\caption{The jet power as a function of both broad line luminosity and Eddington luminosity of Fermi (top) and non-Fermi (bottom) blazars. The meanings of different symbols are as same as Fig.8.}
\label{sample-fig}
\end{figure}
\subsection{Broad line luminosity vs jet power}
The luminosity in the BLR can be taken as an indication of the accretion power of the source (Celotti et al. 1997). We also present a correlation between the broad line emission and jet power for our sample of blazars. Figure 10 and 11 show the broad line luminosity as a function of jet power for Fermi blazars and non-Fermi blazars, respectively. The results of Pearson's analysis show that there are significant correlations between broad line luminosity and jet power for both Fermi blazars and non-Fermi blazars (Fermi blazars: N=76, P$<$0.0001, r=0.74; non-Fermi blazars: N=72, P$<$0.0001, r=0.41). We use a linear regression to analyze the correlation between broad line luminosity and jet power. We obtain $\rm{\log L_{BLR}}\sim\rm{(1.06\pm0.11)\log P_{cav}}$ for Fermi blazars; $\rm{\log L_{BLR}}\sim\rm{(0.59\pm0.16)\log P_{cav}}$ for non-Fermi blazars (95\% confidence level ).

It is generally believed that the jet formation occurs via either BZ (Blandford \& Znajek 1977) mechanisms and /or either the BP (Blandford \& Payne 1982). AGN jet may be driven by both the accretion process and the spin of central BH (Fanidakis et al. 2011; Zhang et al. 2012,2014). Davis \& Laor (2011) suggested that the BH mass would be also an essential factor for the jet radiation efficiency and jet power. The different relationship between jet power and both accretion and black hole mass may indicate the different dominating jet formation mechanisms. We further investigate the connection between the jet properties and both the accretion and central BH. We use multiple linear regression analysis to get the relationships between jet power and both the Eddington luminosity and the broad line region luminosity for Fermi and non-Fermi blazars with 95\% confidence level and r=0.77, 0.44 (Figs. 12);
\begin{eqnarray}
\rm{\log P_{cav}}=\rm{0.52(\pm0.06)\log L_{BLR}+0.09(\pm0.15)\log L_{Edd}}\\
\nonumber{+17.56(\pm6.73)},
\end{eqnarray}
\begin{eqnarray}
\rm{\log P_{cav}}=\rm{0.39(\pm0.11)\log L_{BLR}-0.14(\pm0.16)\log L_{Edd}}\\
\nonumber{+33.72(\pm6.72)}.
\end{eqnarray}
From Equations (2) and (3), we see that both accretion and black hole mass have contributions to the jet power for non-Fermi blazars. However, the black hole mass does not have significant influence on jet power for Fermi blazars. Ghisellini et al.(2014) have suggested that the jet power may depend on the spinning of black hole but the accretion for Fermi blazars.

\section{DISCUSSIONS AND CONCLUSIONS}
In this paper, we study the difference between Fermi and non-Fermi blazars by using a large sample. Our results are as follows:(i) Compared with non-Fermi blazars, the Fermi blazars have lower redshift, lower black hole mass, lower broad line luminosity, lower core luminosity and lower extended luminosity on the average. However, the Fermi blazars have higher average core-dominance parameter than non-Fermi blazars. (ii) Generally, the extend radio luminosity can be used to indicate the intrinsic jet power. The core-dominance parameter can be used as a indicator of beaming effect. The Fermi and non-Fermi blazars could have differences either in intrinsic jet power, or in inclination angle with respect to the beamed jet emission. However, we find that there have no significant difference in intrinsic jet power for all Fermi and all non-Fermi blazars. This result may be explained as follows: Kharb et al.(2010) have suggested that the extend radio luminosity could be affected by interaction with the environment on kiloparsec-scales. The optical luminosity is likely to be a better measure of intrinsic jet power than extend radio luminosity (e.g., Maraschi et al.2008; Ghisellini et al.2009). This is due to the fact that the optical continuum luminosity is correlated with the emission-line luminosity over 4 orders of magnitude (Yee \& Oke 1978), and the emission-line luminosity is tightly correlated with the total jet kinetic power (Rawlings \& Saunders 1991). We also find that there have no significant difference in core-dominance parameter (inclination angle) for all Fermi and all non-Fermi blazars. This result may be explained as follows: Kharb et al. (2010) found that the ratio of the radio core luminosity to the k-corrected optical luminosity ($\rm{\log R_{\nu}}$=$\rm{\log\frac{L_{core}}{L_{opt}}}$=$\rm{(\log L_{core}+M_{abs}/2.5)-13.7}$) appears to be a better indicator of orientation than traditionally used radio core-dominance parameter ($\rm{R_{c}}$).

Compared with Fermi FSRQs, the non-Fermi FSRQs have significant higher mass, and significant higher accretion luminosity, and significant lower core and lower extend radio emission, but similar accretion luminosity in Eddington units and similar radio core-dominance parameter (which may imply they are seen at similar angles to the jet). However, there is plainly lots of overlap between the Fermi and non-Fermi FSRQs. If using the extend radio luminosity to indicate the intrinsic jet power, the results may suggest that the Fermi FSRQs have stronger jet than non-Fermi FSRQs. There may be a general think that the Fermi blazars may have large black hole mass (Ghisellini et al. 2010). Our results seem to contradict with the idea. Meier (1999) have demonstrated explicitly that it is not necessary to have a relatively massive black hole mass to produce powerful jet. Many authors have suggested that the jet power is tied to the spinning of black hole based on current accretion and jet production theory (e.g., Blandford \& Znajek 1977; Meier 1999; Ghisellini et al.2014). These results suggest that the Fermi FSRQs may have higher spinning than non-Fermi FSRQs. Linford et al.(2012) have suggested that the LAT FSRQs had higher radio flux densities than non-LAT FSRQs. Dermer et al.(1995) have suggested that the radio should be beamed in a different way to the Fermi flux in the FSRQs. So the radio should be a broader beam as its seed photons come from the magnetic field which is isotropic in the jet frame. The external compton fermi flux is from seed photons from the BLR which are isotropic in the observer frame and highly anisotropic in the jet frame. The Fermi FSRQ should be those with smaller jet angle (Dermer et al 1995).

Urry \& Padovani (1995) have suggested that many of the main properties of blazars can be explained by the relativistic jets. However, jets formation remains one of the unsolved fundamental problems in astrophysics (Meier et al. 2001). Many models have been proposed to explain the origin of the jets. Two basic of theoretical models have been suggested for the origin of jets: (1) If the black hole is spinning rapidly, the rotational energy of the black hole is expected to be transferred to the jets by the magnetic fields threading the holes (Blandford \& Znajek 1977), (2) The jet can also be accelerated by the large-scale fields threading the rotating accretion disk (Blandford \& Payne 1982). Our results show that there is significant correlation between jet power and black hole mass for Fermi blazars. The Pearson's analysis show that there are significant correlations between jet power and broad line luminosity for both Fermi blazars and non-Fermi blazars, which support that jet power has a close link with accretion. Many authors have confirmed this result (Rawlings \& Saunders 1991; Falcke \& Biermann 1995; Serjeant et al.1998; Cao \& Jiang 1999; Wang et al. 2004; Liu et al.2006; Xie et al.2007; Ghisellini, et al.2009a, 2009b, 2010, 2011; Gu et al. 2009; Sbarrato et al. 2012). Our results suggest that the jet power depends on both the accretion and black hole mass. A linear regression is applied to analyze the correlation between jet power and broad line luminosity, and we obtain $\rm{\log L_{BLR}\sim(1.06\pm0.06)\log P_{cav}}$ for Fermi blazars; $\rm{\log L_{BLR}\sim(0.59\pm0.16)\log P_{cav}}$ for non-Fermi blazars. Ghisellini et al. (2014) also got a close connection between jet powers and accretion disk luminosity for Fermi blazars. Our result is consistent with them.

Ghisellini (2006) have suggested if relativistic jets are powered by a Poynting flux, the Blandford \& Znajek (1997) power can be written as
\begin{equation}
\rm{L_{BZ,jet}}\sim\rm{(\frac{\alpha}{m})^{2}\frac{R_{S}^{3}}{HR^{2}}\frac{\varepsilon_{B}}{\eta}\frac{L_{disk}}{\beta_{r}}}.
\end{equation}
where $\rm{L_{BZ}}$ is the BZ luminosity; $\rm{\frac{\alpha}{m}}$ is the specific black hole angular momentum; and B is the magnetic field in gauss; $\rm{R_{S}}$=$\rm{2GM_{BH}/c^{2}}$ is the Schwarzschild radius; R is the radius; and $\rm{\varepsilon_{B}}$ is the fraction of the available gravitational energy; $\rm{\eta}$ is he accretion efficiency (Xie et al. 2007); $\rm{\beta_{r}}$ is the radial inflow velocity and H is the disk thickness. Ghisellini (2006) also suggested that the maximum BZ jet power can be written as
\begin{equation}
\rm{L_{jet}\sim\frac{L_{disk}}{\eta}}.
\end{equation}
In addition, in view of current theories of accretion disks, the BLR is ionized by a nuclear source (probably radiation from the disk). Maraschi \& Tavecchio (2003) obtained
\begin{equation}
\rm{L_{BLR}=\tau L_{disk}},
\end{equation}
where $\rm{\tau}$ is the fraction of the central emission reprocessed by the BLR, usually assumed to be 0.1. From equations (5) and (6), we have
\begin{equation}
\rm{L_{BLR}\sim\tau\eta L_{jet}}.
\end{equation}
From equation (7), we have
\begin{equation}
\rm{\log L_{BLR}}=\rm{\log L_{jet}+\log\eta+const}.
\end{equation}
Equation (8) shows that the theoretical predicted coefficient of the $\rm{\log L_{BLR}-\log L_{jet}}$ relation is 1. From our results, it is seen that the coefficient of $\rm{\log L_{BLR}\sim\log P_{cav}}$ relation for Fermi blazars is consistent with the theoretical predicted coefficient while is not for non-Fermi blazars.

\section*{Acknowledgments}
We sincerely thank the Chris Done for valuable
comments and suggestions. We are very grateful to the Science
Foundation of Yunnan Province of China(2012FB140,2010CD046). This work is supported by
the National Nature Science Foundation of China (11063004,11163007,U1231203), and the High-Energy Astrophysics Science and Technology Innovation Team of Yunnan Higher School and Yunnan Gravitation Theory Innovation Team (2011c1). This research has made use of the NASA/IPAC Extragalactic Database (NED),that is operated by Jet Propulsion Laboratory, California Institute of Technology, under contract with the National Aeronautics and
Space Administration.

\begin{table*}
\begin{minipage}{250mm}
\caption{The sample of Fermi blazars.}
\begin{tabular}{@{}crcccccccccccccccrl@{}}
\hline\hline
2FGL name & Class & Redshift & $\rm{S_{core}}$ & $\rm{S_{ext}}$ & Ref & $\rm{\log M}$ & Ref & $\rm{\log L_{BLR}}$ & Ref \\
{(1)} & {(2)} & {(3)} & {(4)} & {(5)} & {(6)} & {(7)} & {(8)} & {(9)} & {(10)} \\
\hline
2FGL J0050.6-0929	&	BZB	&	0.2	&	0.57	&	139.7	&	K10	&		&		&		&		\\
2FGL J0108.6+0135	&	BZQ	&	2.099	&	2.81	&	530.6	&	K10	&	8.83	&	Z09	&	46.13,46.98	&	C99,C97	\\
2FGL J0112.1+2245	&	BZB	&	0.265	&	0.36	&	3.9	&	K10	&		&		&		&		\\
2FGL J0116.0-1134	&	BZQ	&	0.67	&		&		&		&	8.57,8.92	&	Sh12	&	44.62,44.88	&	Sh12	\\
2FGL J0120.4-2700	&	BZB	&	0.559	&		&	168	&	CB99	&	9.54	&	X09	&		&		\\
2FGL J0136.9+4751	&	BZQ	&	0.859	&	1.88	&	8.7	&	K10	&	8.73£¬8.3£¬8.3	&	W02£¬Z09£¬C12	&	44.44	&	C99	\\
2FGL J0137.6?2430	&	BZQ	&	0.835	&		&		&		&	9.11£¬9.13	&	L06£¬W02	&	45.34	&	C99	\\
2FGL J0141.5-0928	&	BZB	&	0.733	&		&	50	&	CB99	&	9.84	&	X09	&		&		\\
2FGL J0205.4+3211	&	BZQ	&	1.466	&	0.65	&	11.7	&	K10	&		&		&		&		\\
2FGL J0217.9+0143	&	BZQ	&	1.715	&	0.45	&	71.1	&	K10	&		&		&		&		\\
2FGL J0222.6+4302	&	BZB	&	0.444	&	0.814	&	1052	&	A85	&	8.6	&	C12	&		&		\\
2FGL J0237.8+2846	&	BZQ	&	1.213	&	2.33	&	99.9	&	K10	&	9.22	&	Sh12	&	45.24,45.39,45.9	&	C99,Sh12,C97	\\
2FGL J0238.7+1637	&	BZB	&	0.94	&	1.51	&	25.5	&	K10	&	9,10.22	&	Sb12,C12	&	43.92	&	C99	\\
2FGL J0252.7?2218	&	BZQ	&	1.419	&		&		&		&	9.4	&	Sh12	&	44.73	&	Sh12	\\
2FGL J0259.5+0740	&	BZQ	&	0.893	&	0.552	&	39	&	M93	&		&		&	43.5	&	C99	\\
2FGL J0315.8?1024	&	BZQ	&	1.565	&		&		&		&	8.33	&	Sh12	&	44.67	&	Sh12	\\
2FGL J0339.4-0144	&	BZQ	&	0.852	&	2.92	&	70.3	&	K10	&	8.89,8.98,8.78	&	L06,W02£¬Z09	&	45.23,45	&	C97,L06	\\
2FGL J0405.8-1309	&	BZQ	&	0.571	&	4.33	&	9.1	&	K10	&	9.08,9.07	&	L06,W02	&	45.25	&	L06	\\
2FGL J0423.2-0120	&	BZQ	&	0.916	&	2.91	&	70.2	&	K10	&	9.03£¬8.41	&	W02£¬L06	&	45.59,44.9	&	C97,C99	\\
2FGL J0424.7+0034	&	BZB	&	0.31	&	1.09	&	6.1	&	K10	&		&		&		&		\\
2FGL J0428.6-3756	&	BZB	&	1.11	&		&	86	&	CB99	&	8.6	&	Sb12	&	44.04	&	Sb12	\\
2FGL J0442.7?0017	&	BZQ	&	0.844	&		&		&		&	8.81	&	C12	&	44.81	&	Sh12	\\
2FGL J0449.4-4350	&	BZB	&	0.107	&	0.0993	&	183.2	&	L08	&		&		&		&		\\
2FGL J0453.1-2807	&	BZQ	&	2.559	&		&		&		&		&		&	46.26	&	C99	\\
2FGL J0501.2-0155	&	BZQ	&	2.286	&	1.66	&	148.1	&	K10	&	9.27,8.66	&	Z09,C12	&	45.3	&	C99	\\
2FGL J0530.8+1333	&	BZQ	&	2.06	&	2.24	&	60.1	&	K10	&		&		&		&		\\
2FGL J0532.7+0733	&	BZQ	&	1.254	&	1.54	&	126.8	&	K10	&	8.43	&	Sh12	&	44.86	&	Sh12	\\
2FGL J0538.8-4405	&	BZB	&	0.894	&		&	220	&	CB99	&	8.8,8.33	&	Sb12,L06	&	45.02,44.84	&	C99,Sh12	\\
2FGL J0608.0-0836	&	BZQ	&	0.872	&	1.2	&	123.9	&	K10	&	8.43,8.825	&	Z09,Sh12	&	44.60,45.33	&	C99.Sh12	\\
2FGL J0635.5-7516	&	BZQ	&	0.653	&		&		&		&	9.41	&	W02	&	45.7	&	C99	\\
2FGL J0654.2+4514	&	BZQ	&	0.928	&		&		&		&	8.17	&	Sh12	&	44.26	&	Sh12	\\
2FGL J0654.5+5043	&	BZQ	&	1.253	&		&		&		&	8.79,7.86	&	Sh12	&	43.97	&	Sh12	\\
2FGL J0710.5+5908	&	BZB	&	0.125	&	0.065	&	95	&	GM04	&	8.26	&	W02	&		&		\\
2FGL J0710.8+4733	&	BZB	&	1.292	&	0.973	&	94	&	M93	&		&		&		&		\\
2FGL J0721.9+7120	&	BZB	&	0.3	&	0.69	&	376.4	&	K10	&	8.1	&	C12	&		&		\\
2FGL J0733.9+5023	&	BZQ	&	0.72	&	0.69	&	82.5	&	K10	&	8.84	&	Z09	&		&		\\
2FGL J0738.0+1742	&	BZB	&	0.424	&	1.91	&	20.4	&	K10	&	8.4	&	L03	&		&		\\
2FGL J0739.2+0138	&	BZQ	&	0.189	&	2.34	&	40.9	&	K10	&	8,8.47,7.86	&	W02,C12,L06	&	44.19	&	C99	\\
2FGL J0747.7+4501	&	BZQ	&	0.192	&	0.795	&	33.09	&	C04	&	8.54	&	S11	&	44.34	&	S11	\\
2FGL J0750.6+1230	&	BZQ	&	0.889	&	1.43	&	27	&	K10	&	8.15	&	L06	&	44.95	&	L06	\\
2FGL J0757.1+0957	&	BZB	&	0.266	&	2.07	&	6.7	&	K10	&		&		&		&		\\
2FGL J0808.2-0750	&	BZQ	&	1.837	&	1.58	&	59.8	&	K10	&		&		&		&		\\
2FGL J0809.8+5218	&	BZB	&	0.138	&	0.184	&	4.26	&	C04	&	8.9	&	Z12	&		&		\\
2FGL J0811.4+0149	&	BZB	&	1.148	&	0.46	&	18.2	&	K10	&	8.5	&	Sb12	&	43.62	&	Sb12	\\
2FGL J0816.5+5739	&	BZB	&	0.054	&	0.14	&	31.2	&	C04	&		&		&		&		\\
2FGL J0825.9+0308	&	BZB	&	0.506	&	1.32	&	4.1	&	K10	&	8.83	&	C12	&	43.64,43.37	&	C97,C99	\\
2FGL J0830.5+2407	&	BZQ	&	0.941	&	0.76	&	62.7	&	K10	&	9.01,9.8,8.7	&	Sb12,C12,Sh12	&	44.99,44.97	&	Sb12,Sh12	\\
2FGL J0831.9+0429	&	BZB	&	0.174	&	0.8	&	150.8	&	K10	&	8.8,8.46	&	Sb12,C12	&	42.57	&	Sb12	\\
2FGL J0834.3+4221	&	BZQ	&	0.249	&		&		&		&	9.68	&	S11	&	43.07	&	S11	\\
2FGL J0841.6+7052	&	BZQ	&	2.172	&	3.34	&	73.6	&	K10	&	9.49£¬9.36	&	Z09£¬L06	&	45.91,46.43	&	C97,L06	\\
2FGL J0854.8+2005	&	BZB	&	0.306	&	1.57	&	10.7	&	K10	&	8.8,8.79	&	Sb12,C12	&	43.58,42.83	&	C99,Sb12	\\
2FGL J0903.4+4651	&	BZQ	&	1.465	&	1.645	&	317	&	M93	&	9.25	&	S11	&	45.26	&	S11	\\
2FGL J0909.1+0121	&	BZQ	&	1.025	&	1	&	38	&	K10	&	9.32,8.55,9.14	&	Sb12,L06,Sh12	&	45.1,45.24,45.27	&	C99,Sh12,Sb12	\\
2FGL J0915.8+2932	&	BZB	&	0.101	&	0.222	&	111	&	A85	&		&		&		&		\\
2FGL J0917.0+3900	&	BZQ	&	1.267	&		&		&		&	8.62	&	S11	&	44.8	&	S11	\\
2FGL J0920.9+4441	&	BZQ	&	2.19	&		&		&		&	9.25,9.31,9.29	&	Sb12,C12,Sh12	&	45.85,45.7	&	Sb12,Sh12	\\
2FGL J0921.9+6216	&	BZQ	&	1.446	&	1.11	&	6.4	&	K10	&	8.93	&	Sh12	&	45.05	&	Sh12	\\
2FGL J0927.9-2041	&	BZQ	&	0.347	&		&		&		&	8.46	&	W02	&		&		\\
2FGL J0929.5+5009	&	BZB	&	0.37	&	0.496	&	7.94	&	C04	&		&		&		&		\\
\hline
\end{tabular}
\end{minipage}
\end{table*}
\begin{table*}
\begin{minipage}{250mm}
\contcaption{.}
\begin{tabular}{@{}crcccccccccccccccrl@{}}
\hline\hline
2FGL name & Class & Redshift & $\rm{S_{core}}$ & $\rm{S_{ext}}$ & Ref & $\rm{\log M}$ & Ref & $\rm{\log L_{BLR}}$ & Ref \\
{(1)} & {(2)} & {(3)} & {(4)} & {(5)} & {(6)} & {(7)} & {(8)} & {(9)} & {(10)} \\
\hline
2FGL J0945.9+5751	&	BZB	&	0.229	&	0.069	&	9.01	&	C04	&	8.57,8.77	&	L11	&		&		\\
2FGL J0948.8+4040	&	BZQ	&	1.249	&	1.23	&	95	&	K10	&	8.95	&	S11	&	45.5	&	S11	\\
2FGL J0956.9+2516	&	BZQ	&	0.707	&	0.483	&	19	&	M93	&	9.34,9,8.7,8.47	&	Sb12,W02,L06,Sh12	&	44.92,44.93	&	C99,Sh12	\\
2FGL J0957.7+5522	&	BZQ	&	0.899	&	2.568	&	381	&	M93	&	8.96,7.87,8.07,8.45	&	Sb12,L06,W02,Sh12	&	44.57,44.59	&	Sb12,Sh12	\\
2FGL J0958.6+6533	&	BZB	&	0.368	&		&	34	&	CB99	&	8.53	&	C12	&	42.63	&	C12	\\
2FGL J1012.5+4227	&	BZB	&	0.364	&	0.07	&	10	&	C04	&		&		&		&		\\
2FGL J1015.1+4925	&	BZB	&	0.212	&	0.39	&	12.31	&	C04	&	8.3	&	Z12	&		&		\\
2FGL J1014.1+2306	&	BZQ	&	0.566	&		&		&		&	8.479,8.54	&	S11	&	44.89	&	S11	\\
2FGL J1017.0+3531	&	BZQ	&	1.228	&		&		&		&	9.1	&	Sh12	&	45.34	&	Sh12	\\
2FGL J1019.0+5915	&	BZB	&	2.025	&	0.074	&	10.4	&	C04	&		&		&		&		\\
2FGL J1031.0+5053	&	BZB	&	0.36	&	0.038	&	1.25	&	C04	&		&		&		&		\\
2FGL J1033.2+4117	&	BZQ	&	1.117	&		&		&		&	8.65,8.61	&	Sb12,Sh12	&	44.93,44.92	&	Sb12,Sh12	\\
2FGL J1037.6+5712	&	BZB	&	0.83	&	0.128	&	2.18	&	C04	&		&		&		&		\\
2FGL J1040.7+0614	&	BZQ	&	1.27	&	1.49	&	11	&	K10	&	8.76	&	Z09	&		&		\\
2FGL J1053.6+4928	&	BZB	&	0.14	&	0.052	&	16.6	&	C04	&		&		&		&		\\
2FGL J1058.4+0133	&	BZB	&	0.888	&	2.7	&	230.8	&	K10	&	8.45,8.37	&	Z09,Sh12	&	44.51,44.52	&	C99,Sh12	\\
2FGL J1058.6+5628	&	BZB	&	0.144	&	0.208	&	13.42	&	C04	&	8.54	&	F08	&		&		\\
2FGL J1104.4+3812	&	BZB	&	0.03	&	0.52	&	181	&	A85	&	8.5,8.29,8.22,	&	Sb12,W02,C12	&	41.7	&	Sb12	\\
2FGL J1121.0+4211	&	BZB	&	0.124	&	0.025	&	0.46	&	C04	&		&		&		&		\\
2FGL J1126.6-1856	&	BZQ	&	1.05	&	0.66	&	12.4	&	K10	&		&		&		&		\\
2FGL J1130.3-1448	&	BZQ	&	1.184	&	4.58	&	59.3	&	K10	&	9.18	&	C12	&	45.77	&	C12	\\
2FGL J1136.3+6736	&	BZB	&	0.134	&	0.04	&	10.7	&	C04	&		&		&		&		\\
2FGL J1136.7+7009	&	BZB	&	0.046	&	0.136	&	217.4	&	C04	&	8.21	&	W02	&		&		\\
2FGL J1143.1+6119	&	BZB	&	0.475	&	0.066	&	3.52	&	C04	&		&		&		&		\\
2FGL J1146.8-3812	&	BZB	&	1.048	&		&	10	&	CB99	&	8.5	&	Sb12	&	44.36,44.60	&	C99,Sb12	\\
2FGL J1146.9+4000	&	BZQ	&	1.089	&		&		&		&	8.98,8.93	&	Sb12,Sh12	&	45.07,45.06	&	Sb12,Sh12	\\
2FGL J1150.1+2419	&	BZB	&	0.2	&	0.664	&	25	&	L08	&		&		&		&		\\
2FGL J1151.5+5857	&	BZB	&	0.118	&	0.137	&	55.2	&	C04	&		&		&		&		\\
2FGL J1159.5+2914	&	BZQ	&	0.725	&	1.55	&	196.1	&	K10	&	9.18,8.54,8.375	&	Sb12,L06,Sh12	&	44.71,44.65	&	Sb12,Sh12	\\
2FGL J1203.2+6030	&	BZB	&	0.066	&	0.157	&	87.4	&	C04	&		&		&		&		\\
2FGL J1206.0-2638	&	BZQ	&	0.786	&		&		&		&	8.59,9	&	L06,W02	&	44.07	&	L06	\\
2FGL J1209.6+4121	&	BZB	&	0.377	&	0.397	&	1.18	&	C04	&		&		&		&		\\
2FGL J1217.8+3006	&	BZB	&	0.13	&	0.355	&	189	&	A85	&	8.12	&	W02	&		&		\\
2FGL J1219.7+0201	&	BZQ	&	0.24	&		&		&		&	8.87	&	S11	&	44.83	&	S11	\\
2FGL J1221.3+3010	&	BZB	&	0.182	&	0.067	&	4.3	&	GM04	&	8.6	&	Z12	&		&		\\
2FGL J1221.4+2814	&	BZB	&	0.102	&	2.058	&	2.2	&	A85	&	7.8	&	C12	&	42.25	&	C12	\\
2FGL J1222.4+0413	&	BZQ	&	0.966	&	0.6	&	155.5	&	K10	&	8.24,8.37	&	Sb12,Sh12	&	44.86,44.97	&	Sb12,Sh12	\\
2FGL J1224.9+2122	&	BZQ	&	0.432	&	1.1	&	956.4	&	K10	&	8.87,8.44,8.9	&	Sb12,C12,Sh12	&	45.21,45.16	&	Sb12,Sh12	\\
2FGL J1229.1+0202	&	BZQ	&	0.158	&	34.89	&	17671	&	K10	&	8.9,7.22,8.92	&	Sb12,W02,L06	&	45.54,45.53,45.82	&	C99,Sb12,C97	\\
2FGL J1243.1+3627	&	BZB	&	1.065	&	0.115	&	32.6	&	C04	&		&		&		&		\\
2FGL J1246.7?2546	&	BZQ	&	0.633	&		&		&		&	9.04	&	W02	&		&		\\
2FGL J1248.2+5820	&	BZB	&	0.847	&	0.18	&	4.2	&	C04	&		&		&		&		\\
2FGL J1253.1+5302	&	BZB	&	0.445	&	0.378	&	42.05	&	C04	&		&		&		&		\\
2FGL J1256.1-0547	&	BZQ	&	0.536	&	10.56	&	2095	&	K10	&	8.9,8.43,8.28	&	Sb12,W02,L06	&	44.61,44.38	&	C99,Sb12	\\
2FGL J1309.4+4304	&	BZB	&	0.69	&	0.055	&	2.87	&	C04	&		&		&		&		\\
2FGL J1310.6+3222	&	BZB	&	0.998	&	1.33	&	69.1	&	K10	&	8.8,9.24,8.57	&	Sb12,C12,Sh12	&	45.09,44.92,44.92	&	C99,Sh12,Sb12	\\
2FGL J1317.9+3426	&	BZQ	&	1.056	&		&		&		&	9.29,9.14	&	Sb12	&	45.07,45.09	&	Sh12	\\
2FGL J1326.8+2210	&	BZQ	&	1.4	&	1.14	&	20.4	&	K10	&	9.24,9.25	&	Sb12,Sh12	&	44.90,44.96	&	Sb12,Sh12	\\
2FGL J1337.7-1257	&	BZQ	&	0.539	&	2.07	&	151	&	K10	&	7.98	&	L06	&	44.43,44.18	&	C97,L06	\\
2FGL J1354.7?1047	&	BZQ	&	0.332	&		&		&		&	8.15	&	W02	&		&		\\
2FGL J1419.4+3820	&	BZQ	&	1.82	&	0.52	&	2.5	&	K10	&	8.59,8.68	&	S11	&	45.1	&	S11	\\
2FGL J1420.2+5422	&	BZB	&	0.153	&	1.058	&	18	&	A85	&	8.74	&	W08	&	43.27	&	X07	\\
2FGL J1428.0?4206	&	BZQ	&	1.522	&		&		&		&	9.7	&	L03	&	44.95	&	C99	\\
2FGL J1428.6+4240	&	BZB	&	0.129	&	0.032	&	29.3	&	GM04	&	9.13	&	W02	&		&		\\
2FGL J1439.2+3932	&	BZB	&	0.344	&		&		&		&	8.95	&	W08	&		&		\\
2FGL J1442.7+1159	&	BZB	&	0.163	&	0.06	&	8.5	&	GM04	&		&		&		&		\\
2FGL J1504.3+1029	&	BZQ	&	1.839	&	1.82	&	38.3	&	K10	&	9.64,8.74,8.94	&	Sb12,L06,Sh12	&	45.30,45.17	&	Sb12,Sh12	\\
2FGL J1510.9?0545	&	BZQ	&	1.185	&		&		&		&	8.97	&	C12	&	45.52	&	C12	\\
2FGL J1512.2+0201	&	BZQ	&	0.219	&		&		&		&	8.84,7.99	&	Sb12,W02	&	43.02	&	Sb12	\\
2FGL J1512.8-0906	&	BZQ	&	0.36	&	1.45	&	180.2	&	K10	&	8.6,8.65,8.2	&	Sb12,W02,L06	&	44.75	&	C99	\\
\hline
\end{tabular}
\end{minipage}
\end{table*}
\begin{table*}
\begin{minipage}{250mm}
\contcaption{.}
\begin{tabular}{@{}crcccccccccccccccrl@{}}
\hline\hline
2FGL name & Class & Redshift & $\rm{S_{core}}$ & $\rm{S_{ext}}$ & Ref & $\rm{\log M}$ & Ref & $\rm{\log L_{BLR}}$ & Ref \\
{(1)} & {(2)} & {(3)} & {(4)} & {(5)} & {(6)} & {(7)} & {(8)} & {(9)} & {(10)} \\
\hline
2FGL J1516.9+1925	&	BZB	&	1.07	&	0.255	&	1.7	&	A85	&		&		&		&		\\
2FGL J1517.7-2421	&	BZB	&	0.049	&	2.562	&	32	&	A85	&	8.1	&	W02	&	41.75	&	X07	\\
2FGL J1540.4+1438	&	BZB	&	0.605	&	1.67	&	71.4	&	K10	&	8.94	&	W04	&	43.36,43.05	&	C97,W04	\\
2FGL J1542.9+6129	&	BZB	&	0.117	&	0.126	&	3.7	&	C04	&		&		&		&		\\
2FGL J1549.5+0237	&	BZQ	&	0.414	&	1.15	&	18.8	&	K10	&	8.61,8.72,8.47,8.67	&	Sb12,W02,L06,Sh12	&	44.67,44.83,44.91	&	C99,Sh12,Sb12	\\
2FGL J1550.7+0526	&	BZQ	&	1.422	&	2.21	&	42.9	&	K10	&	9.38,8.98	&	Sb12,Sh12	&	45.06,45.08	&	Sb12,Sh12	\\
2FGL J1553.5+1255	&	BZQ	&	1.29	&		&		&		&	9.1,8.64	&	Sh12	&	45.20,45.18	&	Sh12	\\
2FGL J1559.0+5627	&	BZB	&	0.3	&	0.181	&	19.7	&	C04	&		&		&		&		\\
2FGL J1607.0+1552	&	BZB	&	0.497	&		&		&		&	8.58	&	F08	&		&		\\
2FGL J1608.5+1029	&	BZQ	&	1.226	&	1.35	&	26.5	&	K10	&	8.64,9.5,8.77	&	Sb12,C12,Sh12	&	45.01,45.07	&	Sb12,Sh12	\\
2FGL J1613.4+3409	&	BZQ	&	1.399	&	2.83	&	20.6	&	K10	&	9.12,9.57,9.6,9.08	&	Sb12,W02,L06,Sh12	&	45.87,45.46,45.50	&	C99,Sh12,Sb12	\\
2FGL J1635.2+3810	&	BZQ	&	1.813	&	2.17	&	32	&	K10	&	9.53,9.67,9.075	&	Sb12,C12,Sh12	&	45.82,45.67,45.76	&	C99,Sh12,Sb12	\\
2FGL J1637.7+4714	&	BZQ	&	0.74	&		&		&		&	8.61,8.52	&	Sh12	&	44.58	&	Sh12	\\
2FGL J1640.7+3945	&	BZQ	&	1.66	&	1.17	&	27.6	&	K10	&	8.5	&	D03	&	45.88	&	C99	\\
2FGL J1653.9+3945	&	BZB	&	0.0337	&	1.376	&	67	&	A85	&	9,9.21	&	Sb12,W02	&	42.2	&	Sb12	\\
2FGL J1719.3+1744	&	BZB	&	0.137	&	0.661	&	11	&	A85	&		&		&		&		\\
2FGL J1725.2+5853	&	BZB	&	0.297	&	0.052	&	20.4	&	C04	&		&		&		&		\\
2FGL J1728.2+0429	&	BZQ	&	0.296	&		&		&		&	8.07,7.72	&	W02,L06	&	44.07	&	C99	\\
2FGL J1727.1+4531	&	BZQ	&	0.717	&	1	&	55.3	&	K10	&	8.22	&	W02	&		&		\\
2FGL J1728.2+5015	&	BZB	&	0.055	&	0.175	&	50	&	A85	&	7.86	&	W02	&		&		\\
2FGL J1731.3+3718	&	BZB	&	0.204	&	0.062	&	40.09	&	C04	&		&		&		&		\\
2FGL J1733.1-1307	&	BZQ	&	0.902	&	6.13	&	517.8	&	K10	&	9.3	&	C12	&	44.83	&	C12	\\
2FGL J1740.2+5212	&	BZQ	&	1.375	&	1.61	&	27.6	&	K10	&	9.32	&	L06	&	45.16	&	L06	\\
2FGL J1742.1+5948	&	BZB	&	0.4	&	0.106	&	5	&	C04	&		&		&		&		\\
2FGL J1749.1+4323	&	BZB	&	0.215	&	0.235	&	19.5	&	C04	&		&		&		&		\\
2FGL J1751.5+0938	&	BZB	&	0.322	&	1.05	&	4.9	&	K10	&	8.7,8.34	&	Sb12,C12	&	43.7	&	Sb12	\\
2FGL J1748.8+7006	&	BZB	&	0.77	&		&	12	&	CB99	&		&		&	44.87	&	WS02	\\
2FGL J1801.7+4405	&	BZQ	&	0.663	&	0.5	&	246.6	&	K10	&		&		&		&		\\
2FGL J1800.5+7829	&	BZB	&	0.684	&	1.98	&	20.8	&	K10	&	8.6,7.92	&	Sb12,L06	&	44.85	&	Sb12	\\
2FGL J1806.7+6948	&	BZB	&	0.051	&	1.2	&	368.7	&	K10	&	8.7,8.51	&	Sb12,W02	&	42	&	Sb12	\\
2FGL J1824.0+5650	&	BZB	&	0.664	&	0.95	&	137.4	&	K10	&	9.26	&	C12	&	43.32	&	C12	\\
2FGL J1838.7+4759	&	BZB	&	0.3	&	0.051	&	1.2	&	C04	&		&		&		&		\\
2FGL J1849.4+6706	&	BZQ	&	0.657	&	0.47	&	101	&	K10	&	9.14	&	W02	&	46.2	&	C99	\\
2FGL J2000.8-1751	&	BZQ	&	0.65	&	1.82	&	9.4	&	K10	&		&		&	44.42	&	C97	\\
2FGL J2000.0+6509	&	BZB	&	0.047	&	0.2	&	60	&	GM04	&	8.09	&	W02	&		&		\\
2FGL J2004.5+7754	&	BZB	&	0.342	&	0.823	&	28.9	&	M93	&	8.8	&	G11	&	43.48	&	X07	\\
2FGL J2025.6-0736	&	BZQ	&	1.388	&		&		&		&	9.7	&	L03	&		&		\\
2FGL J2035.4+1058	&	BZQ	&	0.601	&	0.781	&	40	&	A85	&	7.74,8.26	&	Sh12	&	44.17	&	Sh12	\\
2FGL J2133.8-0154	&	BZQ	&	1.285	&	1.37	&	151.9	&	K10	&		&		&	43.66	&	C12	\\
2FGL J2143.5+1743	&	BZQ	&	0.211	&		&		&		&	8.6,8.74	&	Sb12,W02	&	44.26	&	Sb12	\\
2FGL J2147.3+0930	&	BZQ	&	1.113	&	0.698	&	82	&	M93	&		&		&		&		\\
2FGL J2148.2+0659	&	BZQ	&	0.99	&	2.87	&	27.7	&	K10	&	8.87	&	L06	&	46.24,45.77	&	C97,C99	\\
2FGL J2157.9-1501	&	BZQ	&	0.672	&	2.7	&	304.7	&	K10	&	7.59	&	W02	&	43.68	&	C99	\\
2FGL J2158.8-3013	&	BZB	&	0.117	&	0.252	&	132	&	A85	&	8.7	&	Z12	&		&		\\
2FGL J2202.8+4216	&	BZB	&	0.069	&	1.99	&	14.2	&	K10	&	8.7,8.23	&	Sb12,W02	&	42.52	&	Sb12	\\
2FGL J2204.6+0442	&	BZB	&	0.027	&	0.179	&	656	&	GM04	&	8.1	&	W02	&		&		\\
2FGL J2203.4+1726	&	BZQ	&	1.075	&	0.87	&	74.6	&	K10	&		&		&		&		\\
2FGL J2211.9+2355	&	BZQ	&	1.125	&	0.43	&	0.9	&	K10	&	8.46	&	Sh12	&	44.78	&	Sh12	\\
2FGL J2225.6-0454	&	BZQ	&	1.404	&	7.13	&	91.6	&	K10	&	8.81,8.54	&	W04,C12	&	45.6	&	C99	\\
2FGL J2229.7-0832	&	BZQ	&	1.56	&	0.93	&	8.4	&	K10	&	8.95,8.62	&	Sb12,Sh12	&	45.66,45.45	&	Sb12,Sh12	\\
2FGL J2232.4+1143	&	BZQ	&	1.037	&	6.99	&	148	&	K10	&	8.7,8.64	&	Sb12,C12	&	45.87,45.62	&	C99,Sb12	\\
2FGL J2236.4+2828	&	BZQ	&	0.795	&	1.118	&	3.4	&	M93	&	8.35	&	Sh12	&	44.37	&	Sh12	\\
2FGL J2243.2-2540	&	BZB	&	0.774	&		&		&		&	8.6	&	Sb12	&	43.5,43.46	&	C99,Sb12	\\
2FGL J2253.9+1609	&	BZQ	&	0.859	&	14.09	&	822	&	K10	&	8.7,9.17,8.83	&	Sb12,W02,L06	&	45.65,45.52	&	C99,Sb12	\\
2FGL J2258.0-2759	&	BZQ	&	0.927	&		&		&		&	8.92,9.16	&	L06,W02	&	45.84	&	L06	\\
2FGL J2334.3+0734	&	BZQ	&	0.401	&	0.61	&	38.4	&	K10	&	8.37	&	Sh12	&	44.93	&	Sh12	\\
2FGL J2347.9-1629	&	BZQ	&	0.576	&	1.99	&	142.7	&	K10	&	8.72,8.47	&	W02,L06	&	44.62,44.36	&	C97,C99	\\
2FGL J2359.0-3037	&	BZB	&	0.165	&	0.039	&	27.2	&	GM04	&	8.6	&	W02	&		&		\\
\hline
\end{tabular}
\end{minipage}
\end{table*}

\begin{table*}
\centering
\begin{minipage}{145mm}
\caption{The sample of non-Fermi blazars.}
\begin{tabular}{@{}crcccccccccccccccrl@{}}
\hline\hline
Name & Class & Redshift & $\rm{S_{core}}$ & $\rm{S_{ext}}$ & Ref & $\rm{\log M}$ & Ref & $\rm{\log L_{BLR}}$ & Ref \\
{(1)} & {(2)} & {(3)} & {(4)} & {(5)} & {(6)} & {(7)} & {(8)} & {(9)} & {(10)} \\
\hline
PKS 0003-066	&	BZQ	&	0.347	&	2.66	&	43.9	&	K10	&		&		&	43.12	&	C99	\\
0007+106	&	BZQ	&	0.0893	&	0.08	&	17.6	&	K10	&		&	C12	&	44.14	&	C12	\\
0016+731	&	BZQ	&	1.781	&	0.4	&	7.8	&	K10	&	8.93	&	Z09	&	44.98	&	C12	\\
0119+115	&	BZQ	&	0.57	&	1.24	&	113.7	&	K10	&		&		&		&		\\
1ES 0145+138	&	BZB	&	0.125	&	0.003	&	32.8	&	GM04	&	8.42	&	W02	&		&		\\
0146+056	&	BZQ	&	2.345	&	0.865	&	42	&	M93	&		&		&		&		\\
0149+218	&	BZQ	&	1.32	&	1.089	&	25	&	M93	&		&		&		&		\\
0221+067	&	BZQ	&	0.511	&		&		&		&	7.29	&	W02	&		&		\\
0224+671	&	BZQ	&	0.523	&	1.48	&	149.2	&	K10	&		&		&		&		\\
0229+131	&	BZQ	&	2.059	&	1.118	&	212	&	M93	&		&		&	46.61,45.98	&	C97,C99	\\
IERS B0229+200	&	BZB	&	0.14	&	0.042	&	52.2	&	GM04	&	9.24	&	W02	&		&		\\
1ES 0347-121	&	BZB	&	0.185	&	0.009	&	16.7	&	GM04	&	8.65	&	W02	&		&		\\
0350-371	&	BZB	&	0.165	&	0.023	&	17.1	&	GM04	&	8.82	&	W02	&		&		\\
0400+258	&	BZQ	&	2.109	&	1.382	&	1.3	&	M93	&		&		&	46.58	&	C99	\\
TXS 0446+113	&	BZQ	&	1.375	&	1.56	&	15.4	&	K10	&	9.44	&	S12	&	44.49	&	S12	\\
0548-322	&	BZB	&	0.069	&	0.08	&	218	&	A85	&	8.15	&	Z12	&		&		\\
0607-157	&	BZQ	&	0.323	&	3.02	&	1.1	&	K10	&	9.162	&	G01	&	43.56	&	C12	\\
0642+449	&	BZQ	&	3.396	&	0.65	&	1.4	&	K10	&		&		&	48.66	&	C99	\\
B3 0707+424	&	BZQ	&	1.163	&	0.267	&	33.37	&	C04	&		&		&		&		\\
0711+356	&	BZQ	&	1.62	&	1.543	&	4.2	&	M93	&		&		&	47.06	&	C99	\\
TXS 0724+571	&	BZQ	&	0.426	&	0.437	&	25.64	&	C04	&		&		&		&		\\
IERS B0730+353	&	BZB	&	0.177	&	0.046	&	56.74	&	C04	&		&		&		&		\\
0731+479	&	BZQ	&	0.782	&	0.357	&	47.96	&	C04	&		&		&		&		\\
0738+313	&	BZQ	&	0.632	&	2.16	&	65	&	K10	&	9.57	&	L06	&	45.78	&	L06	\\
0742+103	&	BZQ	&	2.624	&	3.62	&	5.8	&	K10	&		&		&		&		\\
PKS 0745+241	&	BZQ	&	0.409	&	0.719	&	196	&	M93	&	7.92	&	C12	&		&		\\
IERS B0756+503	&	BZQ	&	1.622	&	0.114	&	4.23	&	C04	&	9.52	&	S11	&	45.66	&	S11	\\
0812+367	&	BZQ	&	1.027	&		&		&		&	9.2	&	S11	&	45.29	&	S11	\\
0818-128	&	BZB	&	0.074	&	0.63	&	375	&	A85	&		&		&		&		\\
SBS 0818+506	&	BZQ	&	2.133	&	0.053	&	0.83	&	C04	&	9.71	&	S11	&	45.83	&	S11	\\
PKS 0820+22	&	BZB	&	0.951	&	0.161	&	602.5	&	M93	&	9.5435	&	X09	&	44.16	&	C99	\\
0828+493	&	BZB	&	0.548	&		&	0.025	&	CB99	&	9.01	&	X09	&		&		\\
0833+585	&	BZQ	&	2.101	&	0.678	&	30	&	M93	&	9.8	&	S11	&	45.79	&	S11	\\
0836+182	&	BZB	&	0.28	&	0.31	&	88.3	&	L08	&		&		&		&		\\
0839+187	&	BZQ	&	1.276	&	1.264	&	7	&	M93	&	9.79	&	S11	&	45.73	&	S11	\\
B3 0840+378	&	BZQ	&	1.731	&	0.062	&	4.61	&	C04	&	9.48	&	S11	&	45.42	&	S11	\\
HS 0846+5942	&	BZQ	&	1.71	&	0.012	&	1.51	&	C04	&	9.65	&	S11	&	46.28	&	S11	\\
IERS B0850+536	&	BZQ	&	2.422	&	0.021	&	0.73	&	C04	&	9.95	&	S11	&	45.97	&	S11	\\
0850+581	&	BZQ	&	1.319	&		&		&		&	8.49	&	L06	&	45.66	&	L06	\\
IERS B0850+625	&	BZB	&	0.267	&	0.278	&	9.75	&	C04	&		&		&		&		\\
0923+392	&	BZQ	&	0.695	&	2.83	&	361.8	&	K10	&	9.756,9.4	&	G01,W04	&	46.06,45.23	&	C97,W04	\\
1ES 0927+500	&	BZB	&	0.187	&	0.021	&	1.3	&	GM04	&	8.34	&	W02	&		&		\\
SBS 0949+510	&	BZQ	&	1.09	&	0.105	&	5.51	&	C04	&	9.35	&	S11	&	45.12	&	S11	\\
0955+476	&	BZQ	&	1.882	&	0.62	&	1.1	&	K10	&	9.56	&	S11	&	45.27	&	S11	\\
1022+194	&	BZQ	&	0.828	&		&		&		&	8.9	&	S11	&	45.26	&	S11	\\
IERS B1032+382	&	BZQ	&	1.51	&	0.053	&	1.5	&	C04	&	9.73	&	S11	&	45.57	&	S11	\\
GB6 J1033+4222	&	BZB	&	0.211	&	0.026	&	7.19	&	C04	&		&		&		&		\\
1036+054	&	BZQ	&	0.473	&	0.94	&	57	&	K10	&		&		&		&		\\
1045-188	&	BZQ	&	0.595	&	0.76	&	509.4	&	K10	&	7.308	&	G01	&	43.8	&	C99	\\
GB6 J1049+3737	&	BZQ	&	2.997	&	0.055	&	1.07	&	C04	&		&		&	45.87	&	S11	\\
IERS B1051+391	&	BZQ	&	1.372	&	0.072	&	1.01	&	C04	&	9.54	&	S11	&	45.22	&	S11	\\
1055+201	&	BZQ	&	1.11	&	0.768	&	1826	&	M93	&	9.42	&	S11	&	45.8	&	S11	\\
TXS 1059+599	&	BZQ	&	1.83	&	0.416	&	3.04	&	C04	&		&		&		&		\\
TXS 1108+527	&	BZQ	&	1.285	&	0.04	&	76.17	&	C04	&	9.4	&	S11	&	45.23	&	S11	\\
1116+128	&	BZQ	&	2.126	&	1.888	&	259	&	M93	&	8.88	&	S11	&	45.67	&	S11	\\
SBS 1116+603	&	BZQ	&	2.641	&	0.186	&	5.55	&	C04	&	9.79	&	S11	&	46.36	&	S11	\\
IERS B1121+518	&	BZB	&	0.235	&	0.044	&	5.99	&	C04	&		&		&		&		\\
B3 1128+385	&	BZQ	&	1.74	&	0.869	&	5.4	&	M93	&	9.29	&	C12	&	46.26	&	C12	\\
GB6 J1140+4622	&	BZQ	&	0.114	&	0.079	&	2.72	&	C04	&	8.09	&	S11	&	43.77	&	S11	\\
BWE 1145+5710	&	BZQ	&	0.451	&		&		&		&	8.88	&	S11	&	44.8	&	S11	\\
\hline
\end{tabular}
\end{minipage}
\end{table*}
\begin{table*}
\begin{minipage}{145mm}
\contcaption{.}
\begin{tabular}{@{}crcccccccccccccccrl@{}}
\hline\hline
Name & Class & Redshift & $\rm{S_{core}}$ & $\rm{S_{ext}}$ & Ref & $\rm{\log M}$ & Ref & $\rm{\log L_{BLR}}$ & Ref \\
{(1)} & {(2)} & {(3)} & {(4)} & {(5)} & {(6)} & {(7)} & {(8)} & {(9)} & {(10)} \\
\hline
IERS B1146+531	&	BZQ	&	1.638	&	0.096	&	4.86	&	C04	&	9.51	&	S11	&	45.92	&	S11	\\
SBS 1149+499	&	BZQ	&	1.094	&		&		&		&	9.26	&	S11	&	45.52	&	S11	\\
1150+812	&	BZQ	&	1.25	&	1.89	&	89.2	&	K10	&		&		&		&		\\
1150+497	&	BZQ	&	0.334	&	0.6	&	1100	&	A85	&	8.45	&	L06	&	44.39	&	L06	\\
B3 1159+450	&	BZB	&	0.297	&	0.047	&	59.03	&	C04	&		&		&		&		\\
IERS B1200+483	&	BZQ	&	0.816	&	0.067	&	1.7	&	C04	&	9.47	&	S11	&	45.42	&	S11	\\
IERS B1201+454	&	BZQ	&	1.075	&	0.032	&	1.71	&	C04	&	9.19	&	S11	&	45.38	&	S11	\\
IERS B1212+078	&	BZB	&	0.136	&	0.085	&	65	&	GM04	&	8.99	&	W02	&		&		\\
B3 1212+467	&	BZQ	&	0.72	&	0.161	&	117.41	&	C04	&	9.09	&	S11	&	45.12	&	S11	\\
SBS 1215+521	&	BZQ	&	2.229	&	0.079	&	2.16	&	C04	&	9.14	&	S11	&	46.27	&	S11	\\
SBS 1221+503	&	BZQ	&	1.064	&	0.044	&	1.78	&	C04	&	9.14	&	S11	&	45.18	&	S11	\\
1229+645	&	BZB	&	0.163	&	0.055	&	8	&	GM04	&	9.41	&	W02	&	43.04	&	WS02	\\
TXS 1231+481	&	BZQ	&	0.372	&	0.352	&	10.05	&	C04	&	8.26	&	S11	&	44.8	&	S11	\\
BZQ J1235+5228	&	BZQ	&	1.653	&	0.082	&	1.78	&	C04	&	9.59	&	S11	&	45.5	&	S11	\\
1ES 1255+244	&	BZB	&	0.141	&	0.0065	&	10.1	&	GM04	&	8.58	&	W02	&		&		\\
BZB J1301+4416	&	BZB	&	0.435	&	0.052	&	2.11	&	C04	&		&		&		&		\\
1309+355	&	BZQ	&	0.183	&	0.044	&	0.8	&	C04	&	8.8	&	S11	&	44.51	&	S11	\\
1347+539	&	BZQ	&	0.98	&	0.96	&	87.96	&	C04	&	9.43	&	S11	&	45.31	&	S11	\\
IERS B1354+418	&	BZQ	&	0.697	&	0.023	&	0.84	&	C04	&		&		&		&		\\
BWE 1413+4844	&	BZB	&	0.496	&	0.043	&	0.82	&	C04	&		&		&		&		\\
1354+195	&	BZQ	&	0.72	&	1.309	&	855	&	M93	&	9.44	&	W02	&	46.43	&	C99	\\
1400+162	&	BZB	&	0.244	&	0.233	&	548	&	A85	&		&		&	42.27	&	C97	\\
1402+044	&	BZQ	&	3.209	&		&		&		&	8.94	&	S11	&	46.07	&	S11	\\
B3 1409+429	&	BZQ	&	0.887	&	0.071	&	1.14	&	C04	&	9.51	&	S11	&	45.26	&	S11	\\
IERS B1411+746	&	BZB	&	0.46	&	0.113	&	15	&	C04	&		&		&		&		\\
IERS B1413+487	&	BZB	&	0.496	&	0.043	&	0.82	&	C04	&		&		&		&		\\
SBS 1421+511	&	BZQ	&	0.276	&	0.118	&	21.14	&	C04	&	7.57	&	S11	&	44.17	&	S11	\\
B3 1429+401	&	BZQ	&	1.217	&	0.208	&	2.17	&	C04	&	9.96	&	S11	&	45.82	&	S11	\\
1435+638	&	BZQ	&	2.068	&	0.86	&	22.41	&	C04	&	9.43	&	S11	&	46.05	&	S11	\\
1504-166	&	BZQ	&	0.876	&	2.39	&	11.4	&	K10	&	8.84	&	Z09	&	45.57	&	C99	\\
1522+155	&	BZQ	&	0.628	&	0.335	&	68	&	A85	&	8.47	&	S11	&	44.7	&	S11	\\
SBS 1550+582	&	BZQ	&	1.324	&	0.192	&	8.62	&	C04	&		&		&		&		\\
1602+576	&	BZQ	&	2.85	&	0.333	&	9.2	&	C04	&		&		&		&		\\
B2 1604+27	&	BZQ	&	0.934	&		&		&		&	8.83	&	S11	&	44.91	&	S11	\\
IERS B1612+378	&	BZQ	&	1.53	&	0.049	&	1.85	&	C04	&	9.68	&	S11	&	45.75	&	S11	\\
SBS 1618+530	&	BZQ	&	2.347	&	0.179	&	5.69	&	C04	&		&		&		&		\\
1621+392	&	BZQ	&	1.981	&	0.19	&	2.89	&	C04	&	9.67	&	S11	&	46.01	&	S11	\\
1637+574	&	BZQ	&	0.751	&	1.01	&	71.4	&	K10	&	9.18,9.22	&	W02,L06	&	45.57	&	L06	\\
1641+399	&	BZQ	&	0.593	&	7.95	&	1476.9	&	K10	&	9.03£¬9.27	&	Z09£¬L06	&	45.47	&	L06	\\
B3 1642+458	&	BZB	&	0.225	&	0.105	&	78.97	&	C04	&		&		&		&		\\
1642+690	&	BZQ	&	0.751	&	0.999	&	330	&	M93	&	7.76	&	W02	&	46.15	&	C99	\\
IERS B1649+401	&	BZQ	&	2.342	&	0.043	&	1.17	&	C04	&	9.12	&	S11	&	46.38	&	S11	\\
1655+077	&	BZQ	&	0.621	&	1.27	&	199.1	&	K10	&	7.91	&	Z09	&	43.62	&	C12	\\
1656+053	&	BZQ	&	0.879	&	1.301	&	96	&	M93	&	9.09£¬9.74	&	Z09£¬L06	&	46.26	&	L06	\\
1656+571	&	BZQ	&	1.281	&	0.814	&	158.62	&	C04	&		&		&		&		\\
B3 1659+399	&	BZB	&	0.507	&	0.251	&	3.95	&	C04	&		&		&		&		\\
IERS B1705+717	&	BZB	&	0.35	&	0.0384	&	1.8	&	C04	&		&		&		&		\\
1718+481	&	BZQ	&	1.084	&	0.061	&	1.08	&	C04	&		&		&		&		\\
IERS B1726+552	&	BZQ	&	0.247	&	0.142	&	7.34	&	C04	&		&		&		&		\\
1727+386	&	BZQ	&	1.39	&	0.24	&	5.14	&	C04	&		&		&		&		\\
1741-038	&	BZQ	&	1.054	&	1.7	&	3.5	&	K10	&	9.3	&	L03	&	46.77	&	C99	\\
1751+288	&	BZQ	&	1.115	&	0.27	&	8.1	&	K10	&		&		&		&		\\
1758+388	&	BZQ	&	2.092	&	0.33	&	3.2	&	K10	&		&		&		&		\\
1928+738	&	BZQ	&	0.302	&	3.22	&	356.5	&	K10	&	8.72	&	W04	&	44.61	&	W04	\\
1936-155	&	BZQ	&	1.657	&	1.08	&	10.5	&	K10	&	9.3	&	L03	&		&		\\
2005+403	&	BZQ	&	1.736	&	2.47	&	10.6	&	K10	&		&		&		&		\\
2008-159	&	BZQ	&	1.18	&	0.55	&	7.5	&	K10	&	9.56	&	Z09	&		&		\\
IERS B2021+614	&	BZQ	&	0.227	&	2.67	&	1.3	&	K10	&		&		&		&		\\
2037+511	&	BZQ	&	1.686	&	4.97	&	657.6	&	K10	&		&		&		&		\\
\hline
\end{tabular}
\end{minipage}
\end{table*}
\begin{table*}
\begin{minipage}{145mm}
\contcaption{.}
\begin{tabular}{@{}crcccccccccccccccrl@{}}
\hline\hline
Name & Class & Redshift & $\rm{S_{core}}$ & $\rm{S_{ext}}$ & Ref & $\rm{\log M}$ & Ref & $\rm{\log L_{BLR}}$ & Ref \\
{(1)} & {(2)} & {(3)} & {(4)} & {(5)} & {(6)} & {(7)} & {(8)} & {(9)} & {(10)} \\
\hline
2121+053	&	BZQ	&	1.941	&	1.08	&	4.8	&	K10	&	8.6	&	Z09	&		&		\\
2128-123	&	BZQ	&	0.501	&	1.42	&	39.8	&	K10	&	9.16£¬9.02	&	Z09£¬L06	&	45.42	&	L06	\\
2134+004	&	BZQ	&	1.944	&	4.82	&	6.6	&	K10	&	8.91	&	W04	&	44.75	&	W04	\\
2136+141	&	BZQ	&	2.427	&	1.14	&	0.8	&	K10	&		&		&	45.63	&	C99	\\
2142+110	&	BZQ	&	0.548	&		&		&		&	8.69	&	S11	&	44.84	&	S11	\\
2201+315	&	BZQ	&	0.295	&	1.54	&	378.4	&	K10	&	8.91£¬8.94	&	Z09£¬L06	&	45.46	&	L06	\\
2216-038	&	BZQ	&	0.901	&	1.76	&	312.7	&	K10	&	9.08	&	L06	&	45.79	&	L06	\\
2223+210	&	BZQ	&	1.959	&	1.766	&	47	&	M93	&		&		&	49.817	&	C99	\\
2243-123	&	BZQ	&	0.632	&	2.27	&	27.7	&	K10	&	8.32	&	C12	&	45.28	&	C12	\\
2254+074	&	BZB	&	0.19	&	0.454	&	17	&	A85	&	8.62	&	W02	&	42.82	&	WS02	\\
PKS 2316-423	&	BZB	&	0.054	&	0.25	&	391.9	&	L08	&		&		&		&		\\
2328+107	&	BZQ	&	1.489	&	1.065	&	25	&	M93	&		&		&	48.14	&	C99	\\
2344+092	&	BZQ	&	0.677	&	1.801	&	26	&	M93	&	9.31	&	W02	&	47.73	&	C99	\\
2351+456	&	BZQ	&	1.992	&	2.35	&	6.9	&	K10	&	9.22	&	Z09	&	45.11	&	C97	\\
\hline
\end{tabular}
\end{minipage}
\end{table*}
\bsp
\label{lastpage}
\end{document}